\documentstyle[aps,prb,multicol,graphicx]{revtex}
\title{Phase-coherence and the boson-analogy
of vortex liquids}
\author{A. K. Nguyen and A. Sudb{\o} }
\address{Department of Physics\\
         Norwegian University of Science and Technology,
         N-7034 Trondheim, Norway \\}
\begin{document}
\maketitle
\begin{abstract}
The statistical mechanics of the flux-line lattice in extreme type-II
superconductors is studied within the framework of the uniformly frustrated
anisotropic three-dimensional $XY$-model. It is assumed that the externally
applied magnetic field is low enough to invalidate the lowest Landau-level
approach to the problem. A finite-field counterpart of an Onsager vortex-loop
transition in extreme type-II superconductors renders the vortex liquid
phase-incoherent when the Abrikosov vortex lattice undergoes a first order
melting transition. For the magnetic fields considered in this paper,
corresponding to filling fractions $f$ given by
$1/f=12,14,16,20,25,32,48,64,72,84,96,112$, and $128$, the vortex liquid
phase is not describable as a liquid of well-defined field-induced vortex
lines. This is due to the proliferation of thermally induced
closed vortex-loops with diameters of order the magnetic length in the problem,
resulting in a ``percolation transition" driven by non-field induced vortices
also {\it transverse} to the direction of the applied magnetic field. This
immediately triggers flux-line lattice melting and loss of phase-coherence
along the direction of the magnetic field. Due to this mechanism, the field
induced flux lines loose their line tension in the liquid phase, and cannot be
considered to be directed or well defined. In a non-relativistic
$2D$ boson-analogy picture,
this latter feature would correspond to a vanishing mass of the bosons. Scaling
functions for the specific heat are calculated in zero and finite magnetic
field. From this we conclude that the critical region is of order $10\%$ of
$T_c$ for a mass-anisotropy $\sqrt{M_z/M}=3$, and increases with increasing
mass-anisotropy. The entropy jump at the melting transition is calculated
in two ways as a function of magnetic
field for a mass-ansitropy slightly lower than that in $YBCO$, namely with
and without a $T$-dependent prefactor in the Hamiltonian originating
at the microscopic level and surfacing
in coarse grained theories such as the one considered in this paper.
In the first case, it is found to be $\Delta S = 0.1 k_B$ per pancake-vortex,
roughly independent of the magnetic field for the filling fractions
considered here. In the second case, we find an enhancement of $\Delta S$
by a factor which is less than $2$, increasing slightly with decreasing
magnetic field.  This is still lower than experimental values of
$\Delta S \approx 0.4 k_B$ found experimentally for $YBCO$ using calorimetric
methods. We attribute this to the slightly lower mass-anisotropy used in
our simulations. \\[0.2cm]
Pacs-numbers: 74.20.De, 74.25.Dw, 74.25.Ha,74.60.Ec
\end{abstract}
\noindent

\begin{multicols}{2}

\section{Introduction}
The physics of vortex matter represents a new field of research, which has
opened up after the discovery of large fluctuation effects in the extreme
type-II high-$T_c$ superconductors. In particular, the work of Gammel {\it al.}
\cite{Gammel:L87} and Nelson \cite{Nelson:B89} were important milestones in
the field, suggesting for the first time that the Abrikosov vortex lattice
might melt {\it well below} the zero-field critical temperature. This extension
of ideas originally proposed by Eilenberger \cite{Eilenberger:PR67}, has proved
to be fruitful. The melting of the flux-line lattice (FLL) well below the upper
critical field crossover line due to large anisotropy combined with extreme
nonlocality of vortex-vortex  interactions \cite{Houghton:B89}, as well as its
first order character, are now well established both on theoretical and
experimental grounds. Nonetheless, the nature of the molten phase remains
tantalizingly elusive. Obviously, there exists a $2D$ boson analogy picture
of the low-temperature phase of the FLL, where the corresponding boson system
is an insulating one. This is nothing but Abrikosov's mean-field solution to
the problem \cite{Abrikosov:JETP57}. There is, however,
mounting analytical and numerical evidence that a similar, intuitively
appealing,
picture of the molten phase via a $2D$ {\it non-relativistic} superfluid boson
system \cite{Nelson:B89}, may need a substantial revision
\cite{Tesanovic:B95,Nguyen:L96,Tachiki:L97,Nguyen:B98,Hagenaars:B97}.
As far as experimental results are concerned, the situation  also appears quite
intriguing \cite{Zeldov:N95}. \\
At issue here is to what extent the well-defined flux lines of the
low-temperature FLL phase retain their integrity in the high-temperature
molten phase.
%where
%for
%instance flux-line cutting and recombination are known to be substantial in
%layered
%superconductors \cite{Sudbo:L91,Nguyen:B98}.
In this paper, we address the issue of the character of the vortex liquid
via extensive Monte-Carlo simulations of the uniformly frustrated
three-dimensional $XY$ ($3DXY$) model. \\
An issue of fundamental importance is whether or not a vortex liquid (the
molten phase of the FLL) is a superconductor or not when the vortex system is
not pinned. It is clear that the superfluid response to a current applied
transversely to the field induced vortices is zero at all temperatures,
provided that pinning is absent. In the perfect FLL there will however be a
superfluid response to a current applied parallel to the magnetic field. There
remains the possibility that a finite superfluid response might remain even
in the liquid
phase for this geometry, and that it may vanish only deep in the vortex liquid
as suggested originally by Feigel'man {\it et al.} \cite{Feigelman:B93}.
A main point of this paper is to show that for the moderate mass-anisotropy and
large
range of magnetic fields considered here, this in fact does not appear to be
the case.
Moreover, the reason that this is so has important consequences for the
physical picture of the vortex liquid phase. Below, we give a summary of
the main results of this paper. \\
The specific heat is calculated for filling fractions $f=1/12,...,1/128,0$,
and the results are found to be in good agreement with the experimental results
of
Salamon {\it et al.} \cite{Salamon:B93} and more recent experiments of Roulin
{\it et al.} \cite{Junod:C97}, and  Schilling {\it et al.}
\cite{Schilling:N96,Schilling:B96}.
In particular, we find that the near-logarithmic singularity in zero magnetic
field
(the specific heat exponent $\alpha = -0.007$)
is converted to a broad crossover with a peak value suppressed rapidly compared
to the zero field case. This crossover defines, somewhat arbitrarily, the upper
critical magnetic field. The entropy associated with the suppression of the
specific heat at the crossover is partly compensated by the appearance of a
$\delta$-function anomaly in the specific heat at temperatures well below the
zero-field critical temperature. Such a $\delta$-function peak is identified
unambiguously with the first order melting transition of the FLL. The field
dependence of the temperature at which this appears defines the melting line
of the FLL in the $B-T$-phase diagram of the superconductor.  The field
dependendence of the entropy of the transition is found to be
$\Delta S \sim B$, consistent with the suppression of the broad main
crossover peak in the specific
heat. This implies that the entropy per vortex per layer is essentially
field independent in the field range and anisotropy range investigated here.
We emphasize that the anisotropy considered, $\sqrt{M_z/M} = 3$, is moderate
and somewhat smaller than what is found in YBCO. Nonetheless, the qualitative
aspects of our results conform well with those found in YBCO
\cite{Schilling:N96}. \\
We find that in the field regime we have considered, i.e. fields down to
$B \sim 1-5 \rm{T}$, the melting of the vortex lattice is triggered by a
proliferation of thermally induced closed vortex rings of order the magnetic
length of the system. This immediately leads to a ``percolation" of vortex
loops traversing the entire system in any given direction, and in particular
in a direction perpendicular to the applied magnetic field. Hence, flux lines
which are field induced, will traverse the entire system as it weaves its way
from the bottom to the top of the system. In technical terms, in a simulation
one needs to apply periodic boundary conditions at least once in the
$(x,y)$-directions before a flux line starting at the center of the bottom
layer has reached the top layer. {\it Therefore, it does not make sense to view
the vortex liquid phase as a collection of well defined field-induced flux
lines}. The above picture effectively means that the flux-line tension has
vanished in the liquid phase. Within the $2D$ boson analogy picture, an
equivalent statement would be that the boson mass, which is the analog of the
line
tension, has vanished \cite{Tesanovic:PC97}. \\
Scaling functions for the specific heat are calculated, in zero field as well
as in finite field. From the regime where scaling is found in zero field, we
find the width of the critical region to be $|T-T_c|/T_c \approx 0.1$ for a
mass-anisotropy ratio $\Gamma=\sqrt{M_z/M} = 3$. This is considerably wider
than what one would naively obtain using the Ginzburg-criterion. Moreover, this
is slightly wider than what we have found for the isotropic case. The width of
the
critical region therefore increases slightly with mass anisotropy, for the
moderate values of $\Gamma$ we have considered. Vortex loops are expected to be
particularly important for the statistical mechanics of the FLL, provided that
the melting line is found in the proximity of the critical region. From the
obtained melting curve and the width of the critical region, we find that the
melting curve crosses the critical region curve at a field of order $B=1
\rm{T}$
for $\Gamma = 3$. Below this field, vortex loops will completely
dominate the physics at the melting transition.  \\
This paper is organized as follows. In section II we present the model, the
approximations involved and the physical quantities considered, as well as
updating procedure and the parameters used in our Monte Carlo simulations. In
section III we present and discuss detailed results for the filling fraction
$f=1/20$. In section $IV$ we present results in a broad range of filling
fractions
$1/f \in [12,..,128]$. Section V presents the conclusions of this paper.

\section{The Model}
The phenomenological model considered in this paper for the high-$T_c$
cuprates,
is the uniformly frustrated 3 dimensional anisotropic XY ($3DXY$) model on a
lattice,
\cite{Hetzel:L92,Teitel:B93,Teitel:B97,Tachiki:L97,Nguyen:B98},
defined by the Hamiltonian
\begin{equation}
  H(\{\theta({\bf r})\}) = - \sum_{{\bf r},\mu=x,y,z} J_\mu ~
	\cos[\nabla_\mu \theta({\bf r}) - A_\mu({\bf r}) ],
\label{Hamiltonian}
\end{equation}
where $\theta$ is the local phase of the superconducting complex order
parameter and $\nabla$ is a lattice derivative. Furthermore, the coupling
energy along the $\mu$-axis, $J_\mu$, is defined by
\begin{eqnarray*}
  J_x = J_y = \frac{\Phi_0^2 d}{16 \pi^3 \lambda_{ab}^2} \equiv J_\perp,~~~~~
  J_z = \frac{\Phi_0^2 \xi_{ab}^2}{16 \pi^3 \lambda_c^2 d}.
\end{eqnarray*}
Here $\Phi_0$ is the flux quantum, $\xi_{ab}$ is the superconducting
coherence length within the CuO-planes, and $d$ is the distance between two
$CuO$-layers in {\em adjacent unit cells}. Furthermore, $\lambda_{ab}$
and $\lambda_c$ are the magnetic penetration lengths in the CuO planes, and
along the crystals $c$-axis, respectively. In Eq. \ref{Hamiltonian}, $A_\mu$
is related to the {\em quenched} vector potential ${\bf A}_{vp}$ by
\begin{eqnarray*}
        A_\mu({\bf r}) \equiv \frac{2\pi}{\Phi_0}
        \int_{{\bf r}}^{{\bf r} + \hat{e}_\mu}
        d{\bf r}' \cdot \vec{A}_{vp}({\bf r}') ~,
\end{eqnarray*}
where $\hat{e}_\mu$ is the unit vector along the $\mu$-axis.
This $3DXY$ model is dual to the anisotropic London model in the limit of
$(\lambda_{ab},\lambda_c) \rightarrow \infty$ \cite{Carneiro:E92}. This
limit should be taken with the understanding that the coupling energies
$J_\perp$ and $J_z$ are maintained finite. When
$(\lambda_{ab},\lambda_c) \rightarrow \infty$, gauge fluctuations are
completely suppressed, leaving a quenched vector potential and uniform
magnetic induction. Thus, the $3DXY$ model should give an adequate description
of the physics of extreme type-II single crystal superconductors in the field
regime $B_{c1} << B << B_{c2}$. The condition $B_{c1} << B$ ensures that the
contribution to the magnetic induction from individual flux lines overlap
strongly giving a uniform  magnetic induction. The condition $B << B_{c2}$
ensures that details of the internal structure of the vortex cores are not
essential. Moreover, the 3DXY model should give an adequate description of
the physics of extreme type-II single crystal superconductors in zero-magnetic
field  when gauge fluctuations are not important. \\
In this paper we consider simple tetragonal systems with dimensions
$L_x = L_y = L_\perp$ and $L_z$. The coordinate $(x,y,z)$-axes are taken to
be parallel
to the crystal (a,b,c)-axes, respectively. We measure the in-plane length
scales
($x, y, L_x, L_y$) in units of $\xi_{ab}$ and the length scales along the
z-axis
(z, $L_z$) i units of $d$.  Our unit cell is a simple tetragonal system with
dimensions $e_x = e_y = \xi_{ab}$, $e_z = d$. Periodic boundary conditions are
used in all directions throughout.

\subsection{The internal energy and specific heat}
The specific heat per site $C$ is obtained using the standard fluctuation
formula
\begin{eqnarray}
   \frac{C}{k_B} = \frac{1}{L_\perp L_z} \frac{<H^2> - <H>^2}{(k_BT)^2},
\label{Spes}
\end{eqnarray}
where $k_B$ is Boltzmann's constant.  Results for most temperatures  are
checked for consistency by differentiating the results for the internal energy
with respect to temperature. To estimate the latent heat (entropy jump) at a
first order phase transition, we consider the internal energy per site $E$,
\begin{eqnarray}
	E = \frac{1}{L_\perp L_z} <H>.
\label{E}
\end{eqnarray}
This expression holds as long as we do not include any $T$-dependence
in the Hamiltonian. Such a $T$-dependence could conceivably
arise in effective coarse-grained theories such as the GL-theory, as
first pointed out in Ref. \cite{Hu:B97}.
At a first order phase transition there is a discontinuity in the internal
energy per site $\Delta E$ associated with coexistence of the Abrikosov FLL
phase and the vortex liquid. This in turn gives rise to a $\delta$-function
peak in the specific heat \cite{Tachiki:L97,Nguyen:B98}. For the melting
transition of a vortex line lattice, $\Delta E$ is related to the entropy
jump per vortex lines per layer $\Delta S$ by
\begin{eqnarray}
	\frac{\Delta S}{k_B} = \frac{\Delta E}{f k_BT_m},
  \label{DeltaS}
\end{eqnarray}
where $T_m$ is the melting temperature and $f$ is the vortex lines density
defined
below in Eq.~\ref{FilFact}. For consistency, one may also check this result by
extracting the entropy jump at the melting transition from the scaling of the
height of the $\delta$-function anomaly in the specific heat
\cite{Halperin:L81,Tachiki:L97,Nguyen:B98}
\begin{eqnarray}
C = const + \frac{L^3}{4} \Biggr( \frac{\Delta S}{k_B L^3} \Biggl)^2.
\label{delta-spike}
\end{eqnarray}
\subsection{Scaling functions for the specific heat}
An issue of principle importance is whether or not closed thermally induced
vortex loops, i.e. the critical fluctuations, will influence the melting of the
FLL.
Naively, one expects this to be the case provided that the melting temperature
$T_m(B)$ is within the critical region of the zero-field transition. It is
therefore
a matter of interest to establish the width of the critical region of the
anisotropic $3DXY$-model. To this end, we consider finite-size scaling of the
specific heat. The region of data-collapse of the specific heat evaluated at
various system sizes identifies the width of the critical region. It is also
of interest to find the extension of this critical region to finite fields,
i.e.
the width of the crossover region around the upper critical field, and how it
depends on the anisotropy ratio $\Gamma$. If the melting line is located within
this crossover-region or close to it, one expects a vortex-loop ``blowout"
\cite{Nguyen:B98,Nguyen:L96} to dominate the physics at the melting transition,
analagous to what was suggested to happen in superfluid $He^4$ by
Onsager \cite{Onsager:NC49}. This is particularly important at very low
magnetic
fields, like those considered in the experiments of Zeldov {\it et al.}
\cite{Zeldov:N95}. \\
For the zero-field finite-size scaling of the specific heat, we have used cubic
samples $L \times L \times L$ with $L=16,32,48,64,72,96$ to avoid spurious
geometric effects. To investigate the possibility that the width of the
critical
region may depend on anisotropy, we have considered the two cases $\Gamma=1$
and $\Gamma=3$. \\
The finite-size scaling function for the specific heat may in general be
obtained
in standard fashion from the singular part of the free energy as
\cite{Goldenfeld:FP93,Manousakis:B95}
\begin{eqnarray*}
C(t,L)  = L^{\alpha/\nu} ~ \Phi_{\pm} (|t| L^{1/\nu}),
%\label{Cscal1}
\end{eqnarray*}
or equivalently
\begin{eqnarray*}
\frac{C(t,L)}{C(t,\infty)}   = G_{\pm} (|t| L^{1/\nu}),
%\label{Cscal2}
\end{eqnarray*}
where $|t|=|T-T_c|/T_c$, and where $\Phi_{\pm}(|t| L^{1/\nu})$ and
$G_{\pm}(|t| L^{1/\nu})$ are analytic functions of their arguments. Here,
$\alpha$ is the specific heat critical exponent, and $\nu$ is the critical
exponent of the superconducting correlation length, hyperscaling yields
$\alpha = 2 - D \nu$ in a $D$-dimensional system. As discussed in detail in
Ref. \cite{Manousakis:B95}, a more convenient scaling form  for numerical
purposes is given by
\begin{eqnarray}
\frac{C(t,L)-C(0,\infty)}{C(0,L)-C(0,\infty)} = G_{\pm}(|t| L^{1/\nu}).
\label{Spes.scaling0}
\end{eqnarray}
We will use this scaling form to determine the width of the critical region. \\
In the presence of a general field $X$ with scaling dimension $X \sim
\xi^{-\lambda}$,
where $\xi$ is the correlation length $\xi \sim |t|^{-\nu}$, we have
\begin{eqnarray*}
C(t,X) = |t|^{-\alpha} ~ {\cal{G}}_{\pm}(X|t|^{-\lambda \nu}).
\end{eqnarray*}
Subtracting out the zero-field part, and introducing $y \equiv X ~
|t|^{-\lambda \nu}$
and $\Delta C(t,X) \equiv  C(t,X)-C(t,0)$, we find
\begin{eqnarray*}
|t|^{\alpha} ~ \Delta C(t,X) & = & \biggl[
{\cal{G}}_{\pm}(y)-{\cal{G}}_{\pm}(0) \biggr],  \nonumber \\
X^{\alpha/\lambda \nu} ~ \Delta C(t,X) & = &
y^{\alpha/\lambda \nu} ~ \biggl[
{\cal{G}}_{\pm}(y)-{\cal{G}}_{\pm}(0) \biggr] \equiv {\cal{H}}_{\pm}(y).
\end{eqnarray*}
Choosing $X=B$, the induction, implies that the scaling dimension $\lambda=2$
when gauge-fluctuations are suppressed, as is the case in the uniformly
frustrated $3DXY$-model. Under such circumstances, the induction $B$ will
not acquire anomalous scaling. Hence, we obtain
\begin{eqnarray}
B^{\alpha/2 \nu} ~ \Delta C(t,B) & = &
y^{\alpha/2 \nu} ~ \biggl[
{\cal{G}}_{\pm}(y)-{\cal{G}}_{\pm}(0) \biggr] \equiv {\cal{H}}_{\pm}(y),
\label{Spes.scaling}
\end{eqnarray}
where $y=B |t|^{-2 \nu}$.
We will use the above scaling form Eq. \ref{Spes.scaling} to determine the
width
of the crossover region around the upper critical field, as a function of $B$.
Note, however, that this scaling form is not specific to the $3DXY$-model.
By plotting appropriate ratios of temperature- and field-derivatives of
these scaling functions, one may extract directly the critical exponents
of the system as the slopes of the quantitities being plotted, see for
instance the very detailed analysis of this by Schilling {\it et al.}
\cite{Schilling:B96}. It is conceivable that such a procedure would yield
a curve with a kink in it when $t > 0$, as claimed to be observed by Schilling
{\it et al.} This in itself does not invalidate the $3DXY$-scaling of
high-$T_c$ cuprates. Conceivably, it could be due to a crossover from an $XY$
fixed point to another fixed point, possibly with an anomalously large value
of $\nu \approx 1.5$, based on magnetization data. (Note that the specific heat
data of Schilling {\it et al.} in fact show an opposite trend, more consistent
with a crossover to a Gaussian fixed point. This is to be expected if amplitude
fluctuations of the order parameter were to dominate the phase-fluctuations).
For more details, see the discussion below. \\
We note immediately that the above implies that $|t| \sim B^{1/2 \nu}$ for
finite fields, i.e. the width of the crossover region widens as $B$ increases.
Using the estimate $\nu=2/3$ in three dimensions, we have $|t|\sim B^{3/4}$,
implying that the crossover region around the upper critical on the
low-temperature
side  has a positive curvature in the $B-T$ phase-diagram, which
is also true for the melting curve, for which we have $|t|_{M} \sim B^{\eta}$,
with $\eta \sim 2/3$. The widening of the crossover region is of course
consistent with a broadening of the remains of the zero-field
anomaly in the specific heat, to be calculated below. \\
The crossover curve $\tilde B(T<T_c)$ has a more rapid increase as a
function of $T_c-T$ than the melting curve, recall the exponents $3/4$
and $2/3$, respectively. Due to the finite width of the zero-field critical
region, there should then be a field regime where either the melting curve and
the crossover curve intersect, or where the crossover curve is to the left of
the melting curve in the $B-T$  phase-diagram. This depends on the width of the
zero-field critical regime.  Given the size of this regime, $|t| \leq 0.1$,
the former scenario appears to us to be the more likely one, and this is also
what we  find in our simulations. Hence critical fluctuations, i.e. thermally
induced vortex loops, should substantially influence the FLL melting in a
finite regime of magnetic fields. From our simulations, to be presented below,
we
estimate the relevant field regime to be of order $0-1T$ in an extreme type-II
superconductor with $\Gamma=3$.
\subsection{The helicity modulus}
As a probe of global superconducting phase-coherence, we consider various
helicity moduli $\Upsilon_x$, $\Upsilon_y$ and $\Upsilon_z$.
The helicity modulus $\Upsilon_\mu$ along the $\mu$-direction is defined
as the second derivative of the free energy with respect to a global phase
twist along the $\mu$-direction \cite{Teitel:B93}, explicitly we obtain
for the anisotropic uniformly frustrated $3DXY$-model
\begin{eqnarray*}
	\Upsilon_\mu & = & \frac{1}{L_\perp L_z} \left \langle
	\sum_{{\bf r}, \nu=x,y,z}
	J_\nu ~ \cos[\nabla_\nu \theta({\bf r}) - A_\nu({\bf r})]
	(\hat{e}_\nu \cdot \hat{e}_\mu)^2 \right \rangle
\nonumber \\
	& - & \frac{1}{k_B T L_\perp L_z} \left \langle \left [
	\sum_{{\bf r}, \nu=x,y,z} \! \! \! \! \! \! \!
	J_\nu \, \sin[\nabla_\nu \theta({\bf r}) - A_\nu({\bf r})]
	(\hat{e}_\nu \cdot \hat{e}_\mu) \right ]^2 \right \rangle.
\end{eqnarray*}
When $\Upsilon_\mu$ is finite, the system can carry a supercurrent along the
$\mu$-direction. When $\Upsilon_\mu$ vanishes, resistivity along the
$\mu$-direction
becomes finite. In systems with finite applied field along the $z$-axis,
we expect $\Upsilon_x = \Upsilon_y = 0$ for all temperatures in the continuum
limit. In this case, any applied current along  the $xy$-plane will move the
unpinned flux lines and dissipate energy. Discretization introduces a
potentially singular perturbation by introducing an artificial pinning
potential, the effect of which is more serious in a three-dimensional system
than in a two-dimensional one. In the latter case, the effect of the potential
may in principle be entirely avoided by considering low enough filling
fractions \cite{Franz:L94}, whereas this is not possible in three dimensions
in the thermodynamic limit. The size $L_z$ of systems must therefore be
tailored to the filling fraction $f$ in order to avoid
spurious pinning effects. \\
The thus introduced pinning potential will, at a low enough temperature, pin
the flux lines in their positions, and cause $(\Upsilon_x, \Upsilon_y) \neq 0$
up to a depinning temperature $T_d$. To ensure that this artificially
introduced pinning potential caused by the numerical lattice does not affect
the FLL melting transition, we should consider systems with $T_d$ much lower
than all other ``critical'' temperatures of interest. $T_d$ is controlled
mainly by the filling fraction $f$; we have $T_d \to 0$ as $f \to 0$
\cite{Franz:L94,Nguyen:B98}. To adequately mimick the continuum limit of
interest, low enough filling fractions must therefore be considered.

\subsection{The FLL structure function}
To locate the position of the vortex elements we use the following procedure:
The counterclockwise line integral of the gauge-invariant
phase-differences around any plaquette of the
numerical lattice with surface normal along the $\mu$-direction must always
satisfy
\begin{eqnarray*}
	\sum_{C_i} j_\nu({\bf r}) = 2\pi(n_\mu({\bf r}) - f_\mu),
\\
	j_\nu({\bf r}) = \nabla_\nu \theta({\bf r}) - A_\nu({\bf r}).
\end{eqnarray*}
Here, $C_i$ is the closed path traced out by the links surrounding an arbitrary
plaquette, and $\nu$ represents the Cartesian components of the current in the
directions of the links which comprise the closed path $C_i$. Furthermore,
$j_\nu({\bf r})$ is the current on the link between site {\bf r} and site
{\bf r} + $\hat{e}_\nu$, and $n_\mu({\bf r}) = 0, \pm 1$ represents
a vortex segment penetrating the plaquette enclosed by the path $C_i$.
Here, $f_\mu$ is the vortex lines density along the
$\mu$-direction, and is given by
\begin{eqnarray}
        f_{\mu} = \frac{\sum_{\vec{r}}n_{\mu}(\vec{r})}{L_\perp L_z}.
\label{FilFact}
\end{eqnarray}
To probe the structural order of the vortex-system, we consider the inplane
structure function for $n_z$ vortex segments within the same plane
\cite{Nelson:B89},
\begin{eqnarray*}
  S({\bf k}_\perp) = \frac{1}{f^2 L_\perp^2 L_z}
     \left \langle \sum_z \left | \sum_{{\bf r}_\perp} n_z({\bf r}_\perp,z)
     e^{i {\bf k}_\perp \cdot {\bf r}_\perp} \right |^2 \right \rangle.
\end{eqnarray*}
In the FLL phase we expect to see a periodic array of sharp Bragg peaks in
the ${\bf k}_\perp$-plane. In the vortex liquid phase we expect to see
Bragg rings with radius $k_\perp = 2 \pi/a_v$ and $4\pi/a_v$, characteristic
of a liquid. Here, $a_v$ is the average distance between neighboring vortex
lines.

\subsection{Monte Carlo procedure}
The Monte Carlo updating procedure used in this paper is the following. The
numerical lattice is stepped through in a systematic manner. At each site
a change of the local phase of the superconducting condensate is attempted
by a random amount $\Delta \theta \in [-\pi, \pi \rangle$. The  attempt is
accepted or rejected according to the standard Metropolis algorithm.\\
If the accepted phase change causes the current on a link $j_\mu(\vec{r})$ to
fall exceed the range  $j_{\mu}(\vec{r}) \in [-\pi, \pi \rangle$, an amount
$\pm 2 \pi$ is added to the current such that $j_\mu(\vec{r})$ is brought
back into the primary interval $j_{\mu}(\vec{r}) \in [-\pi, \pi \rangle$.
An important point is that this operation can only generate a closed unit
vortex loop around the link where the current is changed, thereby conserving
the net induction of the system. No net vorticity is ever introduced by the
procedure, and the procedure also guarantees that no  vortex line can start
or end within the sample. One sweep refers to $L_\perp^2L_z$ attempts to
change the phase angle. \\
We fix the height of our systems to $L_z=40$ and let $L_\perp$ vary from 40
to 128 depending on the flux-line density under consideration. In  Refs.
\cite{Tachiki:L97,Nguyen:B98}, it was noted that for systems with moderate
anisotropy ($\Gamma \sim 3$), finite
size effects are rather small when the linear dimension of the system and
the total number of flux lines exceed $\sim 40$.
Thus, we believe that finite size effects will not affect the
conclusions in this paper. Likewise, it was observed by the same authors
that finite-size effects were negligible when $L_z$ was increased
beyond $L_z=40$ for the anisotropy considered here, $\Gamma=3$. This has
motivated our choice of $L_z=40$. \\
In this paper we fix the anisotropy parameter $\Gamma$ to
\begin{eqnarray*}
	\Gamma \equiv \sqrt{\frac{J_\perp}{J_z}} =
		\frac{\lambda_z d}{\lambda_{ab} \xi_{ab}} = 3.
\end  {eqnarray*}
in most simulations. Occasionally, comparison is made for the isotropic case
$\Gamma=1$. The magnetic field $B$ is applied along the crystal $c$-axis,
giving a vortex line density $f$,
\begin{eqnarray}
	f_x = f_y = 0,~~~~~f_z \equiv f =  \frac{B \xi_{ab}^2}{\Phi_0}.
\end{eqnarray}
The flux-line densities $f$ considered are:
$1/f = 12 ~(48)$, $14 ~(56)$, $16 ~(48)$, $20 ~(40)$, $25 ~(50)$,
$32 ~(64)$, $48 ~(48)$, $64 ~(64)$, $72 ~(72)$, $84 ~(84)$,
$96 ~(96)$, $112 ~(112)$, $128 ~(128)$, $\infty ~(64)$. The numbers in the
parentheses denote $L_\perp$ for the corresponding vortex line density.
Note that we have chosen, with our gauge, $L_{\perp}$ for each filling
fraction $f$ in such a way that we ensure that an integer number of
magnetic Brillouin-zones will fit on the reciprocal lattice of each system,
enabling us to
use periodic boundary conditions in the $x,y$-directions. As will be observed,
$L_{\perp}$ is an integer number of $1/f$ in each case. \\
The value of $f$ is prescribed by loading the following
phase-difference pattern onto the numerical lattice
(using Landau gauge) a system with $\theta({\bf r})=0$ for all {\bf r},
\begin{eqnarray*}
	A_y(x,y,z) = 2 \pi f x.
\end{eqnarray*}
The system is then heated to a temperature well above any transition/crossover
temperatures of interest, at which point slow cooling is started.  The filling
fraction $f$ is conserved by our Monte Carlo procedure, and the moves are
carried
out on the gauge-invariant phase-differences on each link. Note that we do not
need to {\it assume} any ground state configuration by this procedure.
Extremely
long simulations, typically $4 \times 10^6-6 \times 10^6$ sweeps, are however
required in order to capture the correct physics at the FLL melting transition
and to reveal any $\delta$-function anomalies in the specific heat at the
melting
transition, particularly at low filling fractions.

\section{Results, ${f}=1/20$}
\subsection{FLL melting and phase-coherence}
To identify the possible different phases and  phase transition(s)/crossover(s)
in a system with finite flux-line density, we first concentrate on results
for the system $f=1/20$. {\em Similar results are found in all other finite
flux-line densities considered in this paper}, to be detailed below. We have
measured temperatures in  units where $k_B=1$.

Fig.~\ref{F20.C.S.Yz.Yx} shows the specific heat per site $C$, the
helicity modulus along the applied field direction $\Upsilon_z$, the helicity
modulus perpendicular to the applied field direction $\Upsilon_x$, and the
inplane structure function $S({\bf k}_\perp$), as functions of temperature.
Fig.~\ref{F20.C.S.Yz.Yx} shows that the inplane structure function
$S({\bf k}_\perp = 2\pi/5,\pi/4)$ has a sharp drop from 0.2 to 0 precisely
at $T_m = 0.531J_\perp$, indicating that the FLL melts at $T_m$ in a first
order phase transition. For a more global view, Fig.~\ref{F20.Str2dz}
illustrates
the density plot of $S({\bf k}_\perp)$ for $k_x,k_y \in [-\pi,\pi]$ at four
different temperatures; $T/J_\perp = 0.450, ~0.530, ~0.531, ~0.532$. It is
clearly seen that the periodic array of sharp Bragg-peaks is converted into a
ring precisely at $T_m = 0.531J_\perp$, within a narrow  temperature region of
$\Delta T = 0.001J_\perp$ around $T_m$.

To clarify whether the phase coherence along the direction of the applied
magnetic field is finite in the vortex liquid phase, we consider the helicity
modulus along $z$-axis, $\Upsilon_z$. In Fig.\ref{F20.C.S.Yz.Yx} it is clearly
seen that $\Upsilon_z$ shows a sharp jump from 0.6 to 0 at $T_m$ precisely
where the FLL melts. This shows that the FLL melts directly into
an {\em incoherent vortex liquid} in a first order phase transition.
We will return to this important point later, since it has important
consequences for the physical picture of the vortex liquid phase.
The above result is in complete agreement with the work of Ref.
\cite{Tachiki:L97} using the $3DXY$-model, the work of Ref. \cite{Hu:B97}
using the lowest Landau-level approximation, and earlier work
by us using the $3D$ anisotropic Villain model \cite{Nguyen:B98}. In all
these  works, it was found that longitudinal phase-coherence is lost as soon as
the vortex lattice melts in the thermodynamic limit.
We emphasize that opposite conclusions were
drawn in earlier work by us and others
\cite{Nguyen:L96,Teitel:B93,Ryu:L97}. We believe that this
discrepancy may be due to one or several of the following three
factors: i) In earlier work,
the system size in the $z$-direction may  not have been large
enough, particularly for the isotropic case, ii)
the simulations were not run for a long enough time, and iii) the results
were obtained upon heating only. Our more recent results in Ref.
\cite{Nguyen:B98} and in the present paper are obtained upon heating {\it and}
cooling. \\
A first order phase transition is manifest in the form of a $\delta$-function
anomaly in the specific heat. From   the height of this anomaly
one may deduce the latent heat of the transition. Fig.\ref{F20.C.S.Yz.Yx}
shows that the anomaly occurs at $T_m = 0.531J_\perp$, precisely
where the structure function and the helicity modulus vanish.
These results are in complete agreement with those of
Refs. \cite{Tachiki:L97} obtained on the uniformly
frustrated $3DXY$-model for $f=1/25$, as well as those found
in Ref. \cite{Nguyen:B98} using the uniformly frustrated
$3D$ anisotropic Villain-model for $f=1/32$. Below we consider
filling fractions down to $f=1/128$, finding that these results
still hold.

The latent heat of the first order FLL melting transition at $T_m$
is obtained from the jump in the internal energy shown in Fig.~\ref{F20.E},
using Eq.~\ref{DeltaS}. Here, the entropy jump per vortex line per layer
is estimated to be $\Delta S = 0.1k_B$. To obtain the spike in the specific
heat we must i) find the transition temperature must be located very
accurately, typically to within a part in $10^3$ and ii) increase the
simulation length to at least 6000000 sweeps over the lattice for each
temperature. The extreme length
of the simulations is necessary to allow the system to switch back and forth
between the ordered phase and disordered phase at the phase transition an
adequate number of times, typically at least ten times.

\subsection{Breakdown of the $2D$ boson analogy}
The specific heat has a broad anomaly at $T_{Bc2} \simeq 1.05 J_\perp \gg T_m$,
indicating a crossover. This broad crossover was unambiguously identified in
our previous work as the remains of a zero field Onsager vortex loop
``blow out'' \cite{Nguyen:L96,Nguyen:B98},  destroying superconductivity on
all length scales.
%As shown in our previous  work \cite{Nguyen:B98} the
%incoherent vortex liquid is characterized by a massive amount of flux-line
%intersectioning, and flux-line cutting \cite{Sudbo:L91} and recombination.
%Thus, the flux lines are no
%longer well defined quantities in the incoherent vortex liquid.
However, since
the remains of the zero field vortex loop ``blow out'' takes place first at
$T_{Bc2} \gg T_m$, superconductivity still exists locally in finite domains in
the incoherent vortex liquid phase, giving  strong  diamagnetic fluctuations
\cite{Teitel:B97}. Since the global phase coherence in all directions is
destroyed in the incoherent vortex liquid phase, the superfluid stiffness is
zero in all directions in this phase, and any applied current through the
system will dissipate
energy. Thus, in the incoherent vortex liquid phase the system has both
finite resistivity in all directions, as well as strong diamagnetic
fluctuations. \\
The numerical lattice is a singular perturbation in a three dimensional system,
and one may ask whether the first order FLL melting transition at $T_m$ is
affected by the artificially introduced pinning potential. To address this
issue we consider the helicity modulus along the x-axis, $\Upsilon_x$.
In Fig.~\ref{F20.C.S.Yz.Yx}, it is seen that the helicity modulus along
$x$-axis$\Upsilon_x$ drops to zero already at $T_d = 0.1 J_\perp \ll T_m$.
{}From this we conclude that above $T_d$  the system exhibits a ``floating
solid'' phase \cite{Franz:L94}. Thus the FLL melting transition at $T_m$ is
{\em not} affected by the pinning potential caused by the numerical lattice. \\
Snapshots of the FLL, the incoherent  vortex liquid phase, and  the normal
metal phase  are shown in Fig.~\ref{F20.VLL}, of the system $f=1/20$ for four
temperatures $T/J_\perp = 0.26, ~0.50, ~0.54, ~0.70$. For clarity only a part
of the system; $x,y \in [0:20], z \in [0:40]$, is shown. For the
system $f=1/20$,  we have found $T_m = 0.53J_\perp$ and $T_{Bc2} =
1.05J_\perp$.
For $T = 0.26J_\perp \ll T_m$,  the flux lines form a hexagonal
lattice. Although there are many thermally induced defects
attached to each flux line, they remain well defined entities. For
$T = 0.50 J_\perp \stackrel{<}{\sim} T_m$,  though the flux lines now
fluctuate substantially, they
nonetheless remain intact. So does the FLL, as evidenced by the results
for the structure function, see Figs. \ref{F20.C.S.Yz.Yx} and
\ref{Str.FilFact}. For  a slightly more elevated temperature
$T = 0.54 J_\perp \stackrel{>}{\sim} T_m$,
the FLL has melted. A  key observation is that, immediately upon melting,
the flux lines are no longer well defined entities, there are many
intersections between flux lines, vortex-loops have proliferated, and
there exists at least one way to percolate from one side of the sample
to the opposite side in any direction.
Thus, for any given direction there always exist at least one ``infinite''
long vortex lines perpendicular to it, and any applied current will move
these ``perpendicular'' vortex lines and dissipate energy. Note that in
this picture vortex lines in the incoherent vortex liquid cannot be described
as world lines of $2D$ {\it non-relativistic} bosons \cite{Nelson:B89}.
Thus, one vortex line
in the center of the system will  meander all the way to the boundary surface
(with surface normal perpendicular the applied field direction) and back
as a field induced flux line weaves its way from the bottom to the top of the
system. This corresponds to zero flux-line tension, and a wandering exponent
$\zeta$ of the flux line which is $\zeta \geq 1$ or, equivalently,
zero bosonic mass in the $2D$ boson analogy. \\
Note that the $2D$ quantum boson system we have in mind when referring
to the work of Ref. \cite{Nelson:B89} is a non-relativistic system.
The picture we have in mind for the liquid phase is more akin to a
{\it relativistic} $2D$ quantum boson system, where the proliferation
of vortex loops and overhangs in the flux-lines
correspond to vacuum-fluctuations in the boson system. This connection has
been nicely exposed in Ref. \cite{Kiometzis:FP95}.

\section{Results, $1/{{f}} \in [12,..,128]$}
\subsection{Structure function $S({\bf k}_\perp)$}
We show in Fig.~\ref{Str.FilFact} the inplane structure function
$S({\bf k}_\perp)$ as a function of temperature for several vortex line
densities $f$; \\
$1/f = 12 ({\bf k}_\perp=5\pi /12,5\pi /12)$,
$16 ({\bf k}_\perp=3\pi /8,-\pi /3)$,
$20 ({\bf k}_\perp=2\pi /5,\pi /4)$,
$25 ({\bf k}_\perp=6\pi /25,9\pi /25)$,
$32 ({\bf k}_\perp=5\pi /16,-7\pi /32)$,
$48 ({\bf k}_\perp=\pi /6,\pi /4)$,
$72 ({\bf k}_\perp=\pi /9,-2\pi /9)$,
$96 ({\bf k}_\perp=3\pi /16,5\pi /58)$,
$128 ({\bf k}_\perp=7\pi /64,5\pi /32)$.
For a given vortex line density,
$S({\bf k}_\perp)$ with the corresponding value of ${\bf k}_\perp$ shows a
sharp drop from $\sim 0.2$ to zero defining a field dependent FLL melting
temperature $T_m(f)$. This clearly shows that, for all values of $f$
considered here, the FLL melts in a first order phase transition.
For decreasing f, the transition temperature $T_m(f)$ increase towards
$T_c$ as expected, see Fig.~\ref{Spike.FilFact}.

\subsection{Helicity modulus along the field direction, $\Upsilon_z$}
In Fig.~\ref{Helz.FilFact} the helicity modulus $\Upsilon_z$ along the
direction of the magnetic field is shown as a function of temperature
for the same set of flux-line densities as for the case of the inplane
structure function.  As $f$ is varied, $\Upsilon_z$ shows a sharp drop
towards zero precisely at the corresponding FLL melting temperature
$T_m(f)$. Thus, we may
conclude that for all filling fractions considered, the FLL melts directly into
an incoherent vortex liquid.  The temperature region where the vortex liquid
and the phase coherence along the applied field coexist, found previously
by several authors
\cite{Teitel:B93,Teitel:B97,Nguyen:L96,Feigelman:B93}, is not found
for any flux-line density $f$ considered in this paper. We believe that the
temperature regime where the vortex liquid exist with
phase coherence along the field direction pertains to thin film
geometries, or are otherwise an artifact too short simulations and hysteretic
behavior in the heating/cooling sequence of the vortex system. The
phase-coherent vortex liquid  does not exist in the thermodynamic limit of
an equilibrium system, at least in systems with moderate anisotropy and
moderate magnetic induction.
The possibility of the existence of a very small magnetic field induction
$B_{lower}$ (dependent on the anisotropy $B_{lower}(\Gamma)$) below which
the phase coherence along the field direction can exist in the vortex liquid
is not completely ruled out by this work. \\
Note that $T_m(f)$ is correlated with the temperature where the corresponding
$\Upsilon_z$ start to fall sharply towards zero, not the lowest temperature
where $\Upsilon_z$ vanish. One may question whether it is correct to take the
temperature where $\Upsilon_z$ starts to show a sharp drop as the temperature
where phase coherence along the applied field vanish. For moderate vortex line
densities this poses no problem, since the drop in $\Upsilon_z$ is very sharp.
However, for $f < 1/48$, the transition extend over a small temperature region.
By experience, we know that when the system size and the number of vortex lines
in the system increases, the drop in $\Upsilon_z$ sharpens and the tail in
$\Upsilon_z$ disappear. We believe therefore that this tail is only a finite
size effect.  \\
The main conclusion of the above discussion is that in the thermodynamic
limit, no phase-coherence exists in the vortex liquid phase. This conclusion
is consistent with mounting evidence from numerical simulations
\cite{Tachiki:L97,Nguyen:B98,Koshelev:B97}, obtained however  only in a limited
filling range $1/f \in [25-36]$.  Our present results extend these conclusion
to much lower filling fractions.

\subsection{Specific Heat}
For all filling fractions down to $f=1/32$, we have found a $\delta$-function
anomaly in the specific heat at $T_m(f)$, indicating a first order phase
transition, see Fig~\ref{Spike.FilFact}. For smaller filling fractions
$f \le 1/48$
we find no clear evidence of a spike in the specific heat. Note that in passing
from the system with $f=1/32$ to the system with $f=1/48$, the number of field
induced flux lines is reduced from 128 to 48. We believe that the observed
``none-existence'' of the $\delta$-function anomalies at the FLL melting
temperature in system with very low flux-line densities is attributable to
two factors: 1) for system with $f \le 1/48$ we have to few field induced
vortex lines in our systems, and 2) the contribution to the specific heat from
the field induced flux lines for these filling fractions is too small compared
to the ``spin-wave" and vortex-loop contributions, to be detected by our
simulations. \\

In Fig.~\ref{Spes.FilFact}, we show the specific heat as a function of
temperature
for the same set of corresponding system sizes and flux-line densities as
previously used in calculating the
inplane structure function and the helicity modulus along the direction of the
applied magnetic field. For decreasing $f$ the crossover temperature
$T_{Bc2}(f)$
increases and moves towards the zero-field critical temperature
$T_{Bc2}(f=0)=T_c$. The broad anomaly in the specific heat
sharpens and the maximum height of the cusp increases, evolving
smoothly towards the zero-field specific heat
singularity at $T_c$. $T_{Bc2}$ denotes the crossover temperature at
which the remains of the zero field vortex-loop ``blowout" takes place.
In  a finite magnetic field the vortex -loop ``blowout'' at $T_{Bc2}(f)$
causes only a crossover and the actual
phase transition takes place at a lower temperature, $T_m(f)$, where the FLL
undergoes a first order melting transition triggered by a proliferation of
vortex loops with diameters at least of order the magnetic length in the
problem.\\

Scaling functions for the  specific heat both for zero field and finite field
are shown in Figs. ~ \ref{Spes.Scale0} and \ref{Spes.Scale}. \\
For the zero-field case, using
Eq. \ref{Spes.scaling0}, it is seen from Fig. \ref{Spes.Scale0} that
data-collapse is obtained over a wide region  out to values of the scaling
variable
$|t| L^{1/\nu} > 10$ for $L \geq 32$ on the low-temperature side of $T_c$.
Note also that the width of the scaling regime is slightly larger for
$\Gamma=3$ than for $\Gamma=1$. We expect this trend to persist with
increasing $\Gamma$; in the extremely case where the layers may be considered
completely decoupled, i.e. $\Gamma \to \infty$, the entire low-temperature
regime is known to be critical \cite{Hubermann:L79}. For smaller $L$,
it appears from our simulations that we do not obtain scaling. Using the value
$\nu = 0.669$ \cite{Manousakis:B95}, we find that the width of the
critical region is given by $|t| \approx 0.1$. \\
The critical scaling of the specific heat is also considerably better
above $T_c$ than below. This is due to the fact that vortex loops, i.e.
the critical fluctuations, to a much larger extent dominate the free energy
above
$T_c$ compared to below $T_c$. Below $T_c$ there is a non-singular
contribution to the free energy, and hence specific heat,  due to spin-wave
fluctuations of the local phase of the order parameter. \\

The scaling function of the specific heat in a finite field, given above in
Eq. \ref{Spes.scaling}, is also calculated for filling fractions $f$ given by
$1/f=12,16,20,25,32,48,72,96,128$ with correponding system sizes identical
to those used for the structure function above. The anisotropy is $\Gamma=3$.
The scaled results are good agreement with the works of
Salamon {\it et al.} \cite{Salamon:B93}, Roulin {\it et al.} \cite{Junod:C97},
and Schilling {\it et al.} \cite{Schilling:B96}. Note that while
Ref. \cite{Schilling:B96} makes the point that $3DXY$-scaling does not appear
to describe well the experimental results above $T_c$ in a single-crystal
$YBa_2Cu_3O_7$ in the field-range $0.75-7T$, it does appear to work well below
$T_c$. The reason for this may be that for an optimally doped compound, the
temperature at which a pseudogap opens up may not be much higher than $T_c$
at which phase-coherence is established. The results are therefore likely to
be influenced by amplitude fluctuations above $T_c$. This should not be the
case below $T_c$. Hence, we believe that the field range considered is not
the only issue, but also that one is observing
a crossover from an XY-critical point to a Gaussian critical point when
increasing the temperature above $T_c$, approaching a mean-field like
temperature $T_{MF}$ where preformed pairs start to be dissociate. An obscuring
factor is that the specific heat data and magnetization data of Schilling
{\it et al.} show opposite trends in their deviation from $XY$-scaling. Note
that the analysis of Ref. \cite{Schilling:B96} is {\it not} specific to the
$XY$-model, the scaling forms that are used are quite general. Were the
temperature scales  $T_c$ and $T_{MF}$ to be well separated, $XY$ critical
scaling would presumably persist above $T_c$. This would for instance be the
case in {\it underdoped} cuprates \cite{Kivelson:N95}. \\
At any rate, it is the width of the critical region {\it below} $T_c$ which is
of
interest in establishing the importance of interplay between vortex loops and
FLL melting. The width of the critical region should increase with underdoping,
and hence the interplay between vortex loops and FLL melting is expected to be
more pronounced when the cuprates become more underdoped \cite{Kivelson:N95}.
\\
We note that the scaling is  better above $T_c$ than below, again because
non-singular contributions to the free energy, in this case also arising from
the FLL, contribute significantly. The spikes in the finite-field scaling
function are due to the specific heat anomalies at the FLL melting transition.

\subsection{Entropy discontinuity at the FLL melting transition}
The latent heat, or equivalently the discontinuity in entropy at the FLL
melting transition, has been much focused on in recent experiments
\cite{Safar:L92,Pastoriza:L94,Zeldov:N95,Schilling:N96,Junod:C97}. In
Fig.\ref{Entro.Jump}, the entropy discontinuity at the first order FLL melting
transition is shown as a function of the flux-line density. The results
obtained using the Hamiltonian in Eq. \ref{Hamiltonian} is shown
in filled circles. We find that the
entropy discontinuity per flux line per layer $\Delta S(f) \sim 0.1k_B$.
The fact that $\Delta S$ is essentially independent  of the applied magnetic
field, {\it for the moderate anisotropy $\Gamma = 3$ considered in this paper},
is consistent with the experimental results obtained by Schilling
et al. \cite{Schilling:N96} and Roulin et al. \cite{Junod:C97}.
They found  $\Delta S(B) \sim 0.5k_B$, independent of B.
The values of $\Delta S = 0.1k_B$ are similar to the values found by
Hu et al. \cite{Tachiki:L97}.
We attribute the difference between our values for $\Delta S(f) \sim 0.1k_B$
and the experimental value $\Delta S(B) \sim 0.5k_B$ to the difference in
the anisotropy. YBCO has an anisotropy $\Gamma \sim 7$, while the anisotropy
in this paper is $\Gamma = 3$. As shown in our previous paper
\cite{Nguyen:B98}, and also by Hu et al. \cite{Tachiki:L97}, the entropy
jump at the FLL melting transition increases with increasing anisotropy. \\

To ensure that the artificial pinning potential introduced by the
numerical mesh does not affect the FLL melting transition at $T_m(f)$,
we must ensure that the helicity modulus perpendicular to the applied field
vanishes at a temperature $T_d(f)$ significantly below  $T_m(f)$. Under such
circumstances, the low temperature phase for $T_d(f) < T < T_m(f)$ is
characterized by a ``floating solid phase", mimicking the continuum limit. In
Fig.~\ref{Helx.FilFact}, the helicity modulus along x-direction $\Upsilon_x$
is shown as a function of temperature for the same set of $f$ used for the
specific heat, structure function, and $\Upsilon_z$.
Fig.~\ref{Helx.FilFact} shows that for each flux-line density considered,
$\Upsilon_x$ vanishes at a temperature $T_d(f)$ significantly lower than the
corresponding FLL melting temperature $T_m(f)$. Thus, we have shown that in
all systems considered in this paper, the depinning crossover at $T_d(f)$ does
not affect the FLL melting at $T_m(f) \gg T_d(f)$. Although
we have not shown it explicitly here, we have checked that $\Upsilon_y(T)$ is
essentially  identical to $\Upsilon_x$, as required by symmetry.

In  recent work \cite{Dodgson:L98}, it was pointed out that calculated
entropy jumps $\Delta S$ at the melting transition of the Abrikosov vortex
lattice could be brought into agreement with experiments \cite{Schilling:N96}
by introducing  temperature dependent
parameters in the theory, reflecting  fluctuations at a microscopic
level surfacing in coarse grained theories. The idea of using
such a procedure was first introduced by in Ref. \cite{Hu:B97}
within the lowest Landau level approach to the same problem, i.e.
the high-field limit. This leads to an internal energy
\begin{equation}
U(T)=<H>-T<\frac{\partial H}{\partial T}>,
\end{equation}
where $H$ is an effective
$T$-dependent Hamiltonian, $<H>=(1/Z) \sum ~ H ~ \exp(-H/k_B T)$, and
$Z=\sum \exp(-H/k_B T)$ is the
canonical partition function.  For a derivation of this result,
see Appendix A.

In extreme type-II superconductors, as modelled by
the $3DXY$-model or the London model in the $\lambda \to \infty$-limit, the
$T$-dependence described above appears exclusively as a prefactor in the
Hamiltonian, $H=E_0(\tau) H_0$, where $H_0$ has no $T$-dependent prefactors,
$\tau=T/T_{cMF}$ with $T_{cMF}$ a mean-field zero-field transition temperature,
and $E_0(\tau)=[\lambda(0)/\lambda(\tau)]^2$. $H_0$ is to be identified
with the Hamiltonian used in this paper so far. For instance, in the two-fluid
model $E_0(\tau)=1-\tau^2$, while the simplest mean-field approximation yields
$E_0(\tau)=1-\tau$. Using the above, we find the internal energy given by
\begin{eqnarray}
U       & = &\bigl[E_0(\tau)-T \frac{dE_0(\tau)}{dT} \bigr]~~U_0(T'), \nonumber
\\
U_0(T') & = & \frac{1}{Z} ~~ \sum H_0 \exp(-H_0/k_B T'), \nonumber \\
    T'  & = & \frac{T}{E_0(\tau)}.
\end{eqnarray}
This leads to an entropy
jump at the first-order melting transition of the Abrikosov vortex lattice
\begin{eqnarray}
\Delta S & =  & \frac{\Delta U}{T}
 = \bigl[ E_0(\tau) - T \frac{dE_0(\tau)}{dT} \bigr] ~ \frac{\Delta
U_0(T')}{T} \nonumber \\
 &=&\frac{1}{E_0(\tau)}~~\biggl[E_0(\tau)-T \frac{dE_0(\tau)}{dT}
\biggl]~\Delta S_0(T'),
\end{eqnarray}
where $\Delta S_0(T')= \Delta U_0(T')/T'$ is the entropy jump obtained without
any $T$-dependent parameters in the Hamiltonian, but where the quantity is to
be evaluated at the temperature $T'=T/E_0(\tau)$. Note that the prefactor
relating $\Delta S$ to $\Delta S_0$ would always be $1$ irrespective of what
$E_0(\tau)$ is, if we had not included the contribution
$-T<\partial H/\partial T>$ to $U$. Using $E_0(\tau)=1-\tau^2$, we find
\begin{equation}
\Delta S(T) = \frac{1+\tau^2}{1-\tau^2} ~~~ \Delta S_0(T'),
\label{rdeltas}
\end{equation}
precisely as in Ref. \cite{Dodgson:L98}. Note the difference in the arguments
of $\Delta S(T)$ and $\Delta S_0(T')$. Ref. \cite{Dodgson:L98} concludes that
within a line-liquid model with moderate values of $\Delta S_0$, substantially
enhanced values for $\Delta S$ are obtained, particularly in the low-field
regime, in agreement with experiments. The main factor in the enhancement is
the denominator $1-\tau^2$, which vanishes as $T \to T_{cMF}$.

Note that the above procedure of substituting $H_0$ with $E_0(\tau) ~ H_0$
does not in itself in any way assume that the physics of the vortex system in
the low-field regime is determined exclusively by field-induced flux lines.
However, were we to follow Ref. \cite{Dodgson:L98} and in addition assume that
in the low-field regime, there only exists one relevant length scale in the
problem, namely the magnetic length $a_0 \sim 1/\sqrt{B}$, we {\it would} be
assuming that only field-induced vortices are relevant degrees of freedom
on the melting line. Our main point is that this may be  questionable in
the low-field regime, and we will therefore refrain from utilizing such an
assumption.

If we insist on comparing $\Delta S_0(T')$ with results obtained
using $H_0$, and not $H$ \cite{Dodgson:L98}, then we must fix $T'$ to values
obtained for the melting line in such calculations. {\it Thus, $\tau$ cannot
vary arbitrarily between $0$ and $1$, while fixing $\Delta S_0$ independently.}
Rather, $\tau$ and $T'$ are related via $T' = T/E_0(\tau)$. In calculations of
$\Delta S_0$ using $H_0$, we must therefore have $T'/T_{cMF}<1$. Using
$E_0(\tau) = 1-\tau^2$, we find $\tau<(\sqrt{5}-1)/2$. This gives enhancement
factors $(1+\tau^2)/(1-\tau^2) < \sqrt{5}$ within the two-fluid model.
If we express  Eq. \ref{rdeltas} in terms of $\tau'= T'/T_c$, we obtain
\begin{eqnarray}
\Delta S = \sqrt{1 + 4 \tau'^2} ~~ \Delta S_0(T');~~~ \tau' \in [0,1>.
\label{deltas}
\end{eqnarray}
Similar enhancement factors may be found using the simplest mean-field
approximation $E_0(\tau)=1-\tau$, as for instance used in Ref.
\cite{Koshelev:B97}.
It would yield an enhancement  factor in Eq. \ref{deltas} given by
$1+ \tau'$.
In Fig. \ref{Entro.Jump} we have also plotted the entropy jump as obtained
using a $T$-dependent prefactor in the Hamiltonian. We have used the results
obtained using $H_0$ and enhanced them by the prefactor in Eq. \ref{deltas}.
The inset of the figure shows the enhancement factor on the melting line
obtained in our simulations. It varies quite slowly as a function of $\tau'$
in the  entire interval. Hence, even if we include
the effect of $E_0(\tau)$ on $\Delta S$, we obtain an
essentially field-independent entropy jump in the field regime considered
in Fig. \ref{Entro.Jump}. For specificity, we have chosen
$E_0(\tau)=1-\tau^2$, and ignored the difference between $T_c$ and the
mean-field
critical temperature. We note also in this context that Ref. \cite{Hu:B97}
finds an entropy jump of the magnitude we have found here within the lowest
Landau-level approximation. Furthermore, Ref. \cite{Ryu:L97} finds similar
results using the isotropic $XY$-model with $f=1/6$, in agreement earlier
simulations on the same filling fraction \cite{Hetzel:L92}. Note the large
difference in filling fractions between the present work and the work of
Refs. \cite{Ryu:L97,Hetzel:L92}. For $f=1/6$, commensuration effects due to
the numerical lattice are severe, and could conceivably lead to overestimates
of the magnitude of $\Delta S$. This has been part of the motivation for
pushing the simulations to the low filling fractions used in this paper.

\subsection{B-T Phase Diagram}
To estimate the real magnetic field induction B corresponding to the flux-line
densities considered in this paper, we use Eq.\ref{FilFact} and take
$\xi_{ab} = 12-15$ \AA . With this value of $\xi_{ab}$, we find the magnetic
field corresponding to the smallest flux-line densities considered (f=1/128)
to be approximate $5-7 T$.
In Fig.~\ref{PhaseDiagram} we show the f-T phase diagram originating from
simulations of the XY model. The  flux-line densities $f$ considered are
$1/f = 12, 14, 16, 20, 25, 32, 48, 64, 72, 84, 96, 112, 128$. We see that the
overall behavior of this phase diagram is consistent with the phase diagram in
YBCO measured by Schilling {\it et al.} \cite{Schilling:N96}, and Junod
{\it et al.} \cite{Junod:C97}. The FLL melting line at $T_m(f)$ separates the
superconducting Abrikosov FLL phase from the incoherent vortex liquid phase,
the latter being characterized  by finite resistivity and strong
diamagnetic fluctuations, with simultaneous loss of Bragg peaks in the FLL
structure factor, flux-line integrity, and global phase-coherence in all
directions. The remains of the zero field vortex loop ``blow out'' around
$T_{Bc2}(f)$ destroys phase-coherence on all length scales, and thus separates
the
incoherent vortex liquid phase from the normal metal phase. The melting line
$T_m(f)$ decreases with  decreasing $f$ with a positive curvature. Note also
that
the width of the critical region is large enough to influence the FLL melting
transition over a sizeable field range. This field range is seen to extend up
to $f \approx 1/256$ which we may conservatively estimate to be at least of
order $0-1T$.

\section{Conclusion}
In this paper, we have investigated characteristics of the molten phase of the
Abrikosov flux-line lattice via Monte-Carlo simulations on the
three-dimensional
uniformly frustrated $XY$-model. Bragg-peaks in the static structure factor and
phase-coherence along the direction of the applied magnetic field are both lost
simultaneously, rendering the vortex liquid phase-incoherent. This behavior is
triggered by thermal excitations of closed vortex loops of diameters of the
order of the average distance between flux lines in the low-temperature lattice
phase. On the melting line, this mechanism suffices to produce highly
nontrivial
vortex configurations with appreciable statistical weight {\it on the template
of field induced vortices.} These configurations are characterized by a
``percolation" of closed vortex-loops threading the entire sample in any
direction.
In particular, this is the case for  directions transverse to the direction of
the
applied magnetic field, which is tantamount to a loss of line tension of the
field-induced flux lines. It renders a picture of the molten phase of the
flux-line lattice in terms of a liquid of well-defined, separated, and directed
line-objects, invalid. Equivalently, a picture in terms of world-lines of $2D$
non-relativistic superfluid bosons is invalid in the liquid phase. An effective
theory of the flux-line lattice melting
and the vortex-liquid phase thus appears to present a formidable
challenge involving the solution of a self-consistent coupled theory of
field-induced flux-line objects, and thermally induced closed vortex-loops
\cite{Tesanovic:B95,Nguyen:L96,Tachiki:L97}. This
coupling must evidently render the flux-line tension equal to zero in the
liquid phase. Unfortunately, it is therefore doubtful that the intuitively
appealing physics of directed polymers is particularly relevant for the vortex
liquid phase. \\
Scaling functions for the specific heat are calculated, both in zero and
finite magnetic field. The zero-field results yield a sizeable critical
region $|T-T_c|/T_c \approx 0.1$, corroborating the notion that critical
fluctuations of extreme type-II superconductors, i.e. vortex loops, will
influence such phenomena as flux-line lattice melting over an appreciable
range of magnetic inductions, possibly up to fields of order $1T$ in
moderately anisotropic superconductors. The field range will depend on
mass-anisotropy, since the width of the critical region and the low-field
shape of the melting curve both appear to be influenced by the layeredness
of the superconductor. \\
The finite-field results for the scaling functions for the specific heat,
as well as the obtained phase-diagram for an anisotropy parameter $\Gamma=3$,
are consistent with experiments on the slightly more anisotropic cuprate
high-$T_c$ superconductor YBCO, with $\Gamma \approx 7$. \\
Finally, we note that columnar defects will not be particularly efficient
in enhancing the critical current density in a superconductor where the
FLL melting line is strongly influenced by thermally excited closed vortex
loops. (The influence of columnar defects on the vortex-system
was studied using Monte-Carlo simulations in Ref. \cite{Wallin:PRB93} for
a filling fraction $f=1/2$).
The vortex-loop susceptibility should be sensitive to the phase-stiffness
of the superconductor. The phase-stiffness is in turn largely controlled by the
superfluid density, and therefore also by the charge-carrier density. In order
to avoid the detrimental effects on  transport properties in high-$T_c$
superconductors from a vortex-loop ``blowout", an increase of the
charge-carrier
density appears to be essential.

\section{Acknowledgments}
Support from the Research Council of Norway (Norges Forskningsr{\aa}d)
under Grants No. 110566/410, No. 110569/410, as well as a grant for
computing time under the Program for Super-computing, is gratefully
acknowledged. We thank S.-K. Chin, A. Hansen, J. S. H{\o}ye, A. E. Koshelev,
and Z. Te{\v s}anovi{\'c}
for discussions. J. Amundsen is acknowledged for assistance in optimizing
our computer codes for use on the Cray T3E.

\appendix
\section{Internal energy}
In this appendix, we give a brief derivation of a generalized expression
for the internal energy of a system with an effective $T$-dependent
Hamiltonian. Consider a system in the canonical ensemble. For illustration,
we will consider the well-known $(P,V,T)$-system. Our result
for the internal energy $U$ does not depend on the nature of the
work-term. The system has a statistical distribution function
given by the canonical law
\begin{eqnarray}
\rho = \frac{1}{Z} ~ e^{-\frac{H}{\theta}},
\end{eqnarray}
where the normalization constant $Z$ is the canonical partition function
\begin{eqnarray}
Z = \sum_{\rm{configurations}} ~ e^{-\frac{H}{\theta}},
\end{eqnarray}
where $\theta$ is a parameter of the distribution function which remains
to be determined, such that
\begin{eqnarray}
\sum_{\rm{configurations}} ~ \rho = 1.
\end{eqnarray}
We insist that this normalization is to be maintained if the parameters
$V$ and $\theta$ are varied differentially. The Hamiltonian and hence
the partition function will depend on $V$ through the wall-potential
of the problem. Let us, arbitrarily, write the partition function in the
following way
\begin{eqnarray}
Z = e^{-\frac{\Psi}{\theta}},
\end{eqnarray}
where $\Psi$ is a system-dependent parameter which also remains to be
determined.
When $V \to V + dV$ and $\theta \to \theta + d \theta$, we will
therefore also need to vary $\Psi \to \Psi + d \Psi$ in order to maintain
correct normalization of $\rho$. Hence, we have
\begin{eqnarray}
\sum ~e^{\frac{\Psi-H(V,\theta)}{\theta}} = 1 =
\sum ~ e^{\frac{\Psi+ d \Psi-H(V+dV,\theta + d \theta)}{\theta + d \theta}}.
\end{eqnarray}
Note that we have allowed $H$ to depend on the statistical parameter $\theta$.
Expanding to first order in all differentials, we obtain
\begin{eqnarray}
\sum ~ & e^{\frac{\Psi-H(V,\theta)}{\theta}} & ~ \biggl[ 1
+ \frac{1}{\theta} ~\biggl( d \Psi -\bigl(
\frac{\partial H}{\partial V} dV \nonumber \\
 & + & \frac{\partial H}{\partial \theta} d \theta \bigr)
-\frac{d \theta}{\theta} \bigl(\Psi -H \bigr) \Biggr) \biggr] = 1.
\end{eqnarray}
Since the orginal distribution prior to changing
$V \to V + dV$ and $\theta \to \theta + d \theta$ also was normalized
we obtain the  following constraint on the differentials $dV$, $d \theta$,
and $d \Psi$
\begin{eqnarray}
d \Psi = \frac{d \theta}{\theta} ~\bigl[
\Psi - <H> + \theta ~ \bigl< \frac{\partial H}{\partial \theta} \bigr>
                                   \bigr]
+ \bigl< \frac{\partial H}{\partial V} \bigr> dV.
\end{eqnarray}
Here, $<..>$ denotes a statistical average with respect to the original
distribution function $\exp((\Psi-H)/\theta)$. In order to make the connection
to thermodynamics, we now compare the above with the ``thermodynamic identity"
\begin{eqnarray}
dF = -S ~ dT - P ~ dV  = \frac{dT}{T} (F-U) - P ~ dV,
\end{eqnarray}
where $F=U-TS$ is Helmholz free energy, $S$ is the entropy,  and $U$ is
the internal energy. This comparison yields directly
\begin{eqnarray}
P & = & \bigl< -\frac{\partial H}{\partial V} \bigr>, \nonumber \\
\frac{d \theta}{\theta} & = & \frac{d T}{T} \rightarrow
\theta = k_B T \nonumber, \\
\Psi & = & F  \nonumber \\
U & = & <H> - T \bigl< \frac{\partial H}{\partial T} \bigr>.
\end{eqnarray}
Note that $\Psi$ thus identified is the only choice consistent
with $F = - k_B T \ln Z$. This then fixes $U$. Also, the expression for
$U$ obtained in this fashion is identical to that obtained directly from
the usual relation
\begin{eqnarray}
U = - \frac{\partial \ln Z}{\partial \beta},
\end{eqnarray}
with an assumed $T$-dependent Hamiltonian.

\end{multicols}

\begin{figure}[htbp]
%  \begin{picture}(0,210)(0,0)
%     \put(-30,210)
%         {\includegraphics[angle=270,scale=0.38]
%         {F20.C.S.Yz.Yx.ps}}
%  \end{picture}
{\small FIG. \ref{F20.C.S.Yz.Yx}. Specific heat $C$ per site, inplane
structure factor $S({\bf k}_\perp = 2\pi/5, \pi/4)$, helicity modulus along
z-axis $\Upsilon_z$, and helicity modulus along x-axis $\Upsilon_x$ as
functions of temperature for the system with vortex line density $f=1/20$. The
inplane structure function $S({\bf k}_\perp)$ jumps discontinuously from 0.2
to 0 precisely at $T_m = 0.531J_\perp$ indicating that the FLL melts in a first
order phase transition. At the same temperature, $\Upsilon_z$ also shows a
discontinuity from 0.6 to 0, indicating that {\em the FLL melts directly
into the incoherent vortex liquid} with no global phase coherence along the
applied magnetic field direction. At temperatures above $T_m$ there is no
global phase coherence in any direction. The specific heat also shows a
$\delta$-function anomaly precisely $T_m$. The broad specific heat anomaly
at $T_{Bc2} \sim 1.05 J_\perp$ represents the remains of the zero-field
Onsager vortex loop blowout. Note that for temperatures
$T_m < T < T_{Bc2}$ local superconducting phase coherence still exist.
giving strong diamagnetic fluctuations in the liquid phase.
The FLL depins from the numerical lattice at $T_d \ll T_m$ where
$\Upsilon_x$ vanish. Thus, the FLL melting transition at
$T_m \gg T_d$ is not affected by the numerical lattice.}
\refstepcounter{figure}
\label{F20.C.S.Yz.Yx}
\end{figure}

\begin{figure}[htbp]
%  \begin{picture}(0,260)(0,0)
%     \put(-55,270)
%        {\includegraphics[angle=270,scale=0.44]
%         {F20.Str2dz.ps}}
%  \end{picture}
{\small FIG. \ref{F20.Str2dz}. Intensity plots of the structure function
$S({\bf k}_\perp)$ for various temperatures for the system with flux-line
density $f=1/20$. $k_x \in [-\pi,\pi]$ and $k_y \in [-\pi,\pi]$ is along the
horizontal and the vertical direction, respectively. The brightness in the
plots is a measure of the magnitude of $S({\bf k}_\perp)$.
To enhance features we put all points where  $S({\bf k}_\perp) < 0.01$
(noise level) to black and all points where $S({\bf k}_\perp) > 0.05$ to white.
Precisely at $T_m$, the sharp Bragg-peaks in $S({\bf k}_\perp)$ are converted
into  Bragg-rings, characteristic of a liquid. Thus, the FLL melts into
a vortex liquid within a temperature region of $\Delta T=0.001J_\perp$.}
\refstepcounter{figure}
\label{F20.Str2dz}
\end{figure}

\begin{figure}[htbp]
%  \begin{picture}(0,210)(0,0)
%     \put(-30,210)
%        {\includegraphics[angle=270,scale=0.38]
%         {F20.E.ps}}
%  \end{picture}
{\small FIG. \ref{F20.E}. Internal energy per site $E$ as a function of
temperature for the system with vortex line density $f=1/20$. The data are
obtained from a cooling sequence using 3000000 sweeps per temperature.
The internal energy has a discontinuous jump
at $T_m$ indicating a first order transition from an ordered state (FLL)
to a disordered state (phase-incoherent vortex liquid). This jump in the
internal energy is used to determine the latent heat (entropy jump)
at the FLL melting transition. The jump in $E$ here corresponds to a jump in
the entropy per vortex line per layer $\Delta S(f=1/20) = 0.1k_B$.}
\refstepcounter{figure}
\label{F20.E}
\end{figure}

\begin{figure}[htbp]
%  \begin{picture}(0,220)(0,0)
%     \put(-35,240)
%        {\includegraphics[angle=270,scale=0.38]
%         {F20.FLL.ps}}
%  \end{picture}
{\small FIG. \ref{F20.VLL}. Snapshots of the vortex configuration for the
system with vortex line density $f=1/20$ for four temperatures,
$T/J_\perp=0.26, 0.50, 0.54, 0.70$. For clarification we have shown only a
part of the system; $x,y \in [0:20]$ and $z \in [0:40]$.
For $T = 0.26 J_\perp \ll T_m$, the flux lines form a hexagonal
lattice. Although there are many thermally induced defects
attached to each flux line, they are nonetheless  well defined quantities. For
$T = 0.50 J_\perp \stackrel{<} \sim T_m$, the FLL
is still intact. Although the flux lines now contain many larger defects, they
are still well defined. For $T = 0.54 J_\perp \stackrel{>}{\sim} T_m$, the FLL
has melted. For $T \stackrel{>} \sim 0.54 J_{\perp}$, it is seen
that the flux lines are no longer well defined quantities.
There exists at least one way for a flux -line to thread the system
in any direction. For any vortex-configuration, therefore, there
exists at least one flux line  threading the sample in the direction
perpendicular to the magnetic field. }
\refstepcounter{figure}
\label{F20.VLL}
\end{figure}

\begin{figure}[htbp]
%  \begin{picture}(0,210)(0,0)
%     \put(-30,210)
%        {\includegraphics[angle=270,scale=0.38]
%         {Strz.FilFact.ps}}
%  \end{picture}
{\small FIG. \ref{Str.FilFact}. The inplane structure function
$S({\bf k}_\perp)$ as a function of temperature for several vortex line
densities $f$. For a given $f$, $S({\bf k}_\perp)$ with the corresponding
value of ${\bf k}_\perp$ shows a sharp drop from $\sim 0.2$ to $0$ at a
well-defined FLL metling temperature $T_m(f)$.}
\refstepcounter{figure}
\label{Str.FilFact}
\end{figure}

\begin{figure}
%  \begin{picture}(0,200)(0,0)
%     \put(-10,200)
%        {\includegraphics[angle=270,scale=0.36]
%         {Helz.FilFact.ps}}
%  \end{picture}
{\small FIG. \ref{Helz.FilFact}. The helicity modulus $\Upsilon_z$ along the
field direction as a function of temperature for several flux-line densities
$f$. For all densities, $\Upsilon_z$ shows a sharp drop
towards zero precisely at the corresponding FLL melting temperature $T_m(f)$.
Thus, the FLL melts directly into an incoherent vortex liquid.}
\refstepcounter{figure}
\label{Helz.FilFact}
\end{figure}

\begin{figure}
%  \begin{picture}(0,210)(0,0)
%     \put(-20,210)
%        {\includegraphics[angle=270,scale=0.38]
%         {Spike.FilFact.ps}}
%  \end{picture}
{\small FIG. \ref{Spike.FilFact}. Monte-Carlo results for the specific
heat per site of the
anisotropic $3DXY$-model as a function of temperature for several vortex line
densities $f$. The system sizes depends on filling fraction, as explained in
the text, and $\Gamma=3$. For clarity the $n$th curves is shifted by an amount
$0.2*n$ upwards.
For each $f$ there is a spike at a $f$-dependent critical temperature $T_m(f)$
indicating a first order phase transition.}
\refstepcounter{figure}
\label{Spike.FilFact}
\end{figure}

\begin{figure}
%  \begin{picture}(0,210)(0,0)
%     \put(-20,210)
%        {\includegraphics[angle=270,scale=0.38]
%         {Spes.FilFact.ps}}
%  \end{picture}
{\small FIG. \ref{Spes.FilFact}. Monte-Carlo results for the specific heat
of the anisotropic
$3DXY$-model as a function of temperature for several flux-line
densities. For decreasing $f$ (decreasing magnetic field induction B) the
crossover temperature $T_{Bc2}(f)$ increases and moves towards the zero
field critical temperature $T_{Bc2}(f=0)=T_c$. The broad anomaly
(cusp) in the specific heat becomes sharper and the maximum height of the
cusp increases. Thus, for decreasing $f$ the specific heat
evolves smoothly to the zero field specific heat singularity at $T_c$.
The spike in the specific heat at $T_m(f) \ll T_{Bc2}(f)$ each graph is hidden
in the noise of the other graphs and is therefore hard to recognize in this
particular figure.
While in zero magnetic field the vortex loop blowout is the
mechanism for the second order phase transition at $T_c$, in finite magnetic
field the vortex loop blowout at $T_{Bc2}(f)$ is only a crossover.
The phase transition in systems with finite vortex line densities take place
at a lower temperature, $T_m(f)$, where the vortex line lattice melts.}
\refstepcounter{figure}
\label{Spes.FilFact}
\end{figure}

\begin{figure}[htbp]
%  \begin{picture}(0,365)(0,0)
%     \put(-35,-20)
%        {\includegraphics[scale=0.50]
%         {Finite.Size.ps}}
%  \end{picture}
{\small FIG. \ref{Spes.Scale0}. Monte-Carlo results for the specific heat
of the anisotropic $3DXY$-model in zero magnetic field, for various system
sizes $L \times L \times L$ with $L=32,48,64,72,96$, and two values of the
anisotropy, $\Gamma=1$ and $\Gamma=3$, scaled according to
Eq.~\ref{Spes.scaling0}. Here, $t = (T-T_c)/T_c$. The region of data
collapse gives the width of the critical region. Note that this region is
slightly for $\Gamma=3$ than for $\Gamma=1$.}
\refstepcounter{figure}
\label{Spes.Scale0}
\end{figure}

\begin{figure}
%  \begin{picture}(0,210)(0,0)
%     \put(-20,-140)
%        {\includegraphics[scale=0.45]
%         {B.Scaled.ps}}
%  \end{picture}
{\small FIG. \ref{Spes.Scale}. Monte-Carlo results for the specific heat
of the anisotropic
$3DXY$-model for a number of filling fractions $f$ given by $1/f=12,..,128$
with corresponding system sizes as explained in text, and anisotropy
$\Gamma=3$, scaled according to Eq.~\ref{Spes.scaling}. The results are in
good agreement with the experimental results of Schilling {\it et al.},
and Junod {\it et al.}}
\refstepcounter{figure}
\label{Spes.Scale}
\end{figure}

\begin{figure}
%  \begin{picture}(0,210)(0,0)
%     \put(-50,-150)
%        {\includegraphics[angle=0,scale=0.5]
%         {DeltaS.ps}}
%  \end{picture}
{\small FIG. \ref{Entro.Jump}. The entropy jump per vortex line per layer
$\Delta S(f)$ at the FLL
melting transition for several vortex line densities $f$.
The filled circles represent the results obtained with
a $T$-independent Hamiltonian, Eq. \ref{Hamiltonian}. The
open circles represent the results obtained including a $T$-dependent
prefactor in the Hamiltonian. We see that $\Delta S(f)$ essentially does
not depend on $f$ in this regime of filling fractions $f$, regardless
of whether $T$-dependent prefactors are included in the Hamiltonian
or not. The inset shows the enhancement factor in Eq. \ref{deltas}. }
\refstepcounter{figure}
\label{Entro.Jump}
\end{figure}

\begin{figure}
%  \begin{picture}(0,210)(0,0)
%     \put(-30,210)
%        {\includegraphics[angle=270,scale=0.38]
%         {Helx.FilFact.ps}}
%  \end{picture}
{\small FIG. \ref{Helx.FilFact}. The helicity modulus perpendicular to the
field direction
$\Upsilon_x$ as a function of temperature for several vortex line
densities $f$. For each $f$, $\Upsilon_x$ vanishes at a temperature
$T_d(f)$ significantly lower than the corresponding FLL melting temperature
$T_m(f)$. The artificial  pinning potential of the
numerical lattice therefore does not affect
the FLL melting transition at $T_m(f) \gg T_d(f)$.}
\refstepcounter{figure}
\label{Helx.FilFact}
\end{figure}

\begin{figure}
%  \begin{picture}(0,250)(0,0)
%     \put(-20,-80)
%        {\includegraphics[scale=0.45]
%         {PhaseDiagram.ps}}
%  \end{picture}
{\small FIG. \ref{PhaseDiagram}. The f-T phase diagram for the uniformly
frustrated 3d XY model.
The applied field is along the crystal c-axis, the anisotropy parameter
$\Gamma = 3$. The FLL exhibits global phase coherence along the applied
field direction. The FLL phase is separated from the incoherent flux-line
liquid phase by the melting line $T_m(f)$.
The melting transition is a first order phase transition with an entropy
jump $\Delta S(f) \sim 0.1k_B$ for the anisotropy $\Gamma=3$ and
field regime considered in this paper. In the incoherent vortex liquid phase
$T_m(f) < T < T_{Bc2}(f)$, there is only local, but no global, phase coherence
in any direction. At finite fields, between the incoherent vortex liquid and
the normal metal phase, there exists a broad crossover region where a blowout
of thermally induced closed vortex loops takes place, eventually also
destroying
superconductivity on short length scales. The width of the crossover regime is
obtained from scaling  behavior of the specific heat. Another, consistent,
method of obtaining
this width, is to estimate the temperature regime which correponds
to an uncertainty of $10 \%$ in the maximum value of the specific
heat anomaly at $T_{Bc2}(f)$. Since this anomaly becomes broader with
increasing field, the cross-over region becomes wider. This is also confirmed
from the scaling results for the specific heat.}
\refstepcounter{figure}
\label{PhaseDiagram}
\end{figure}

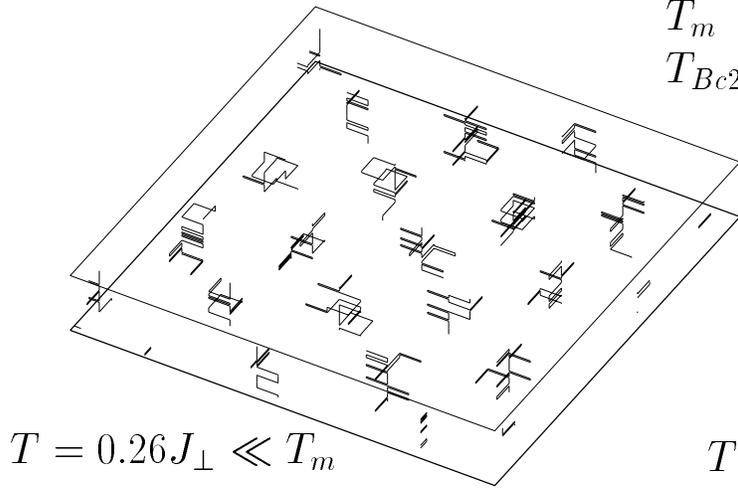
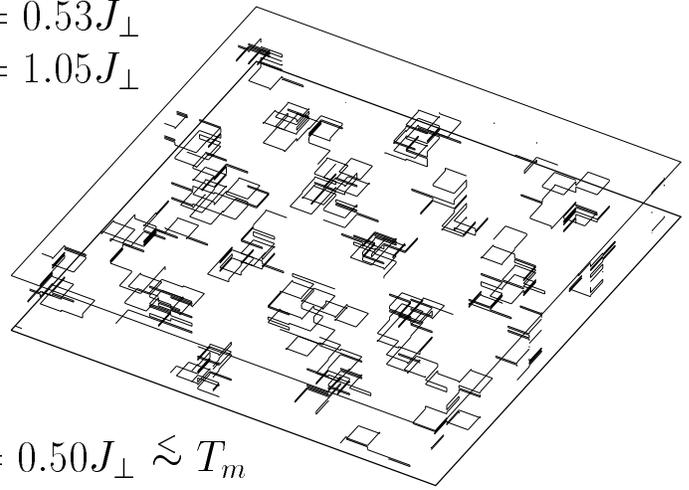
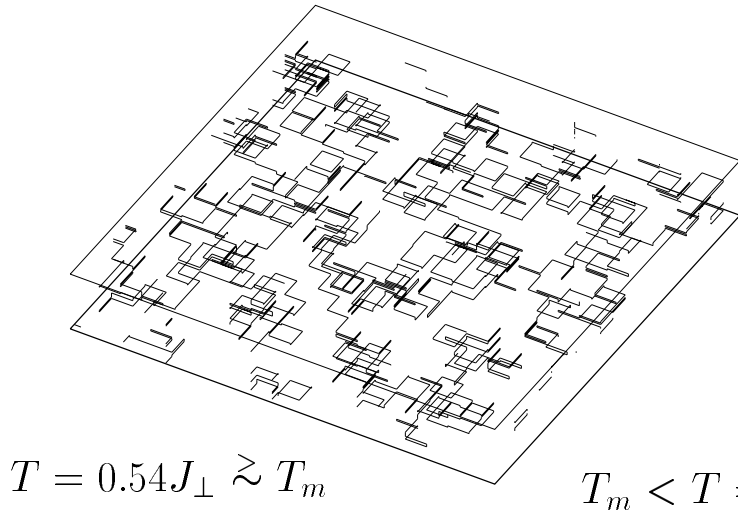
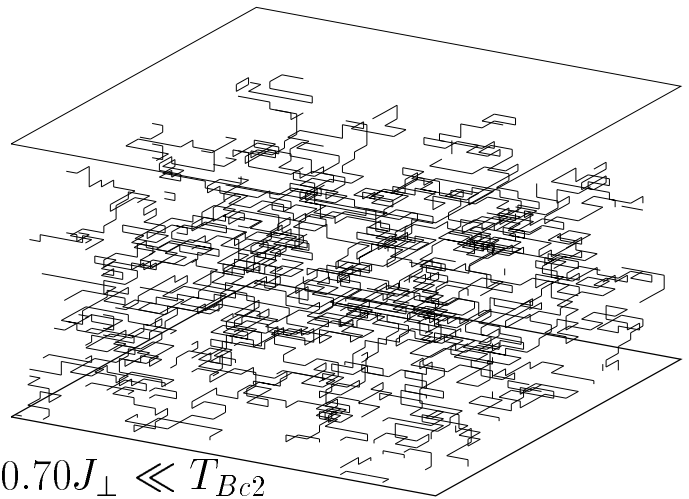
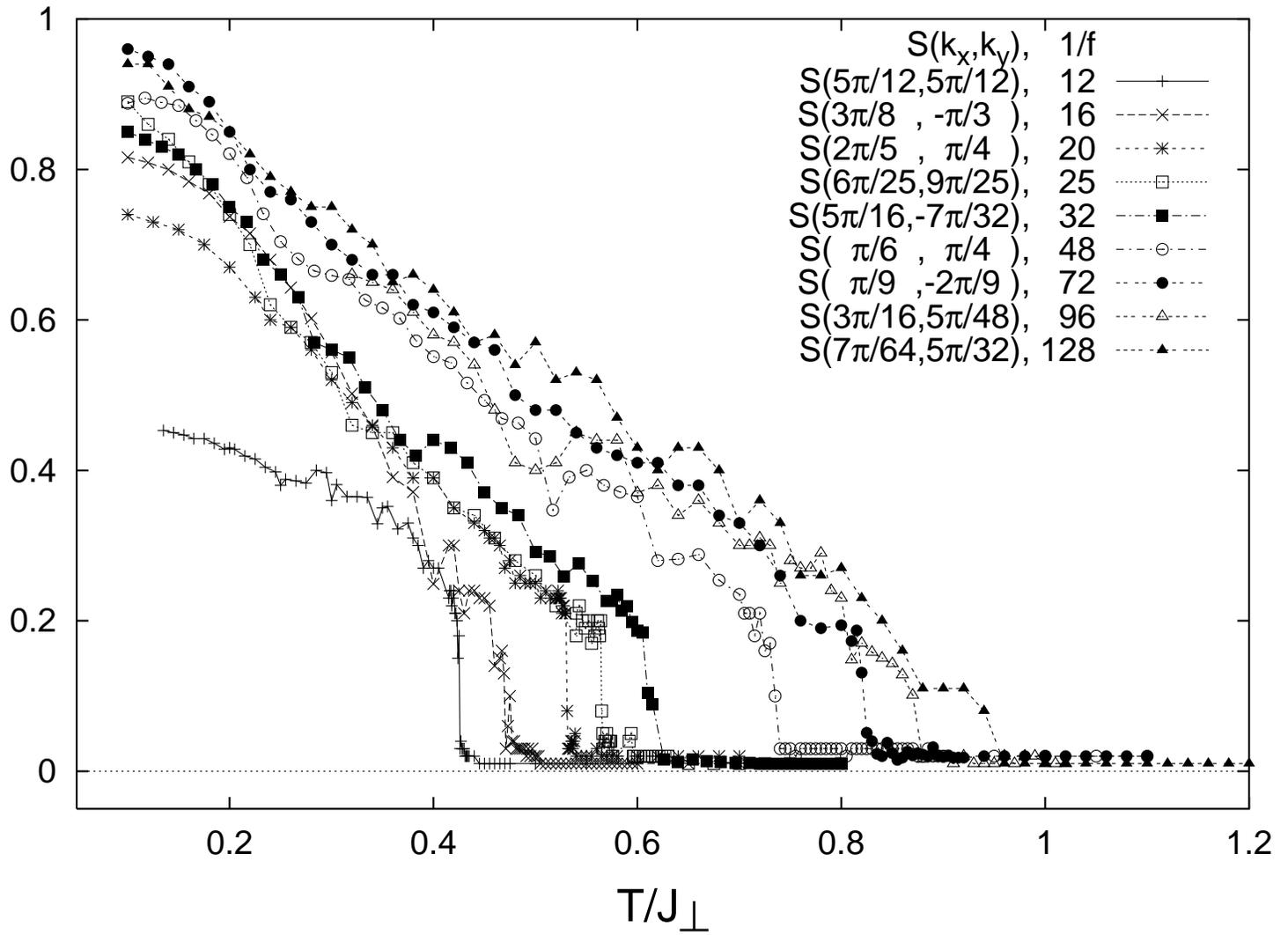
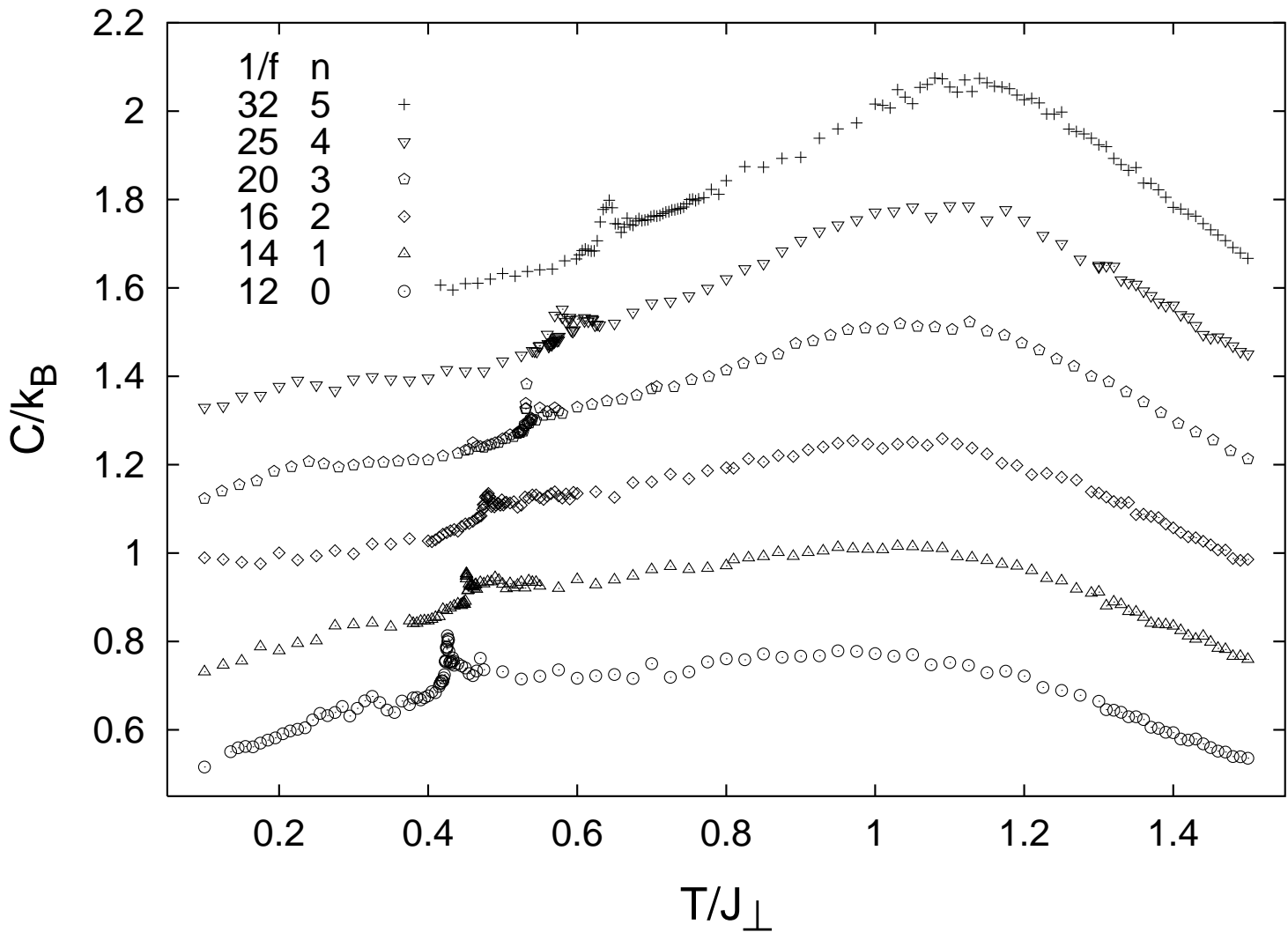
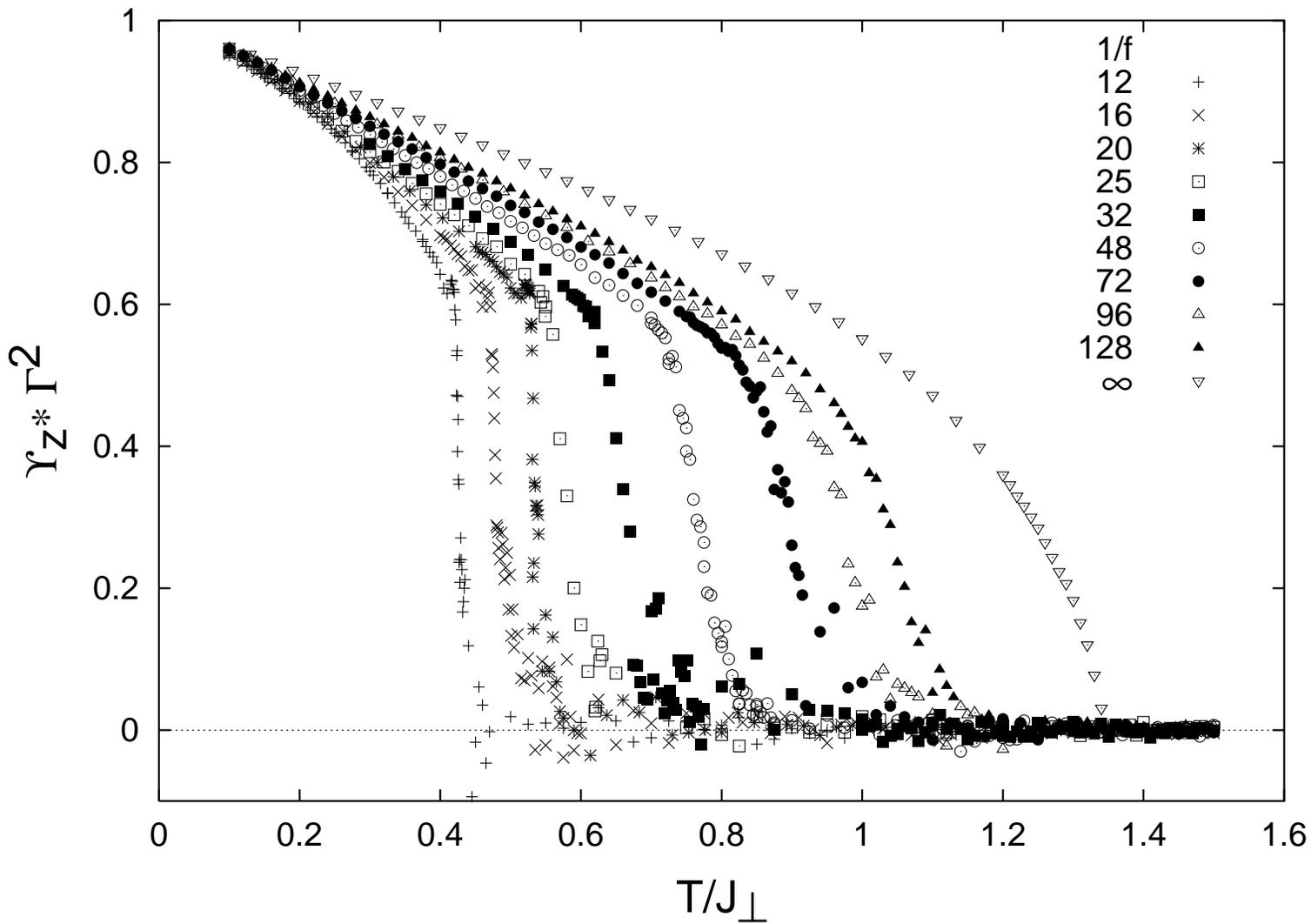
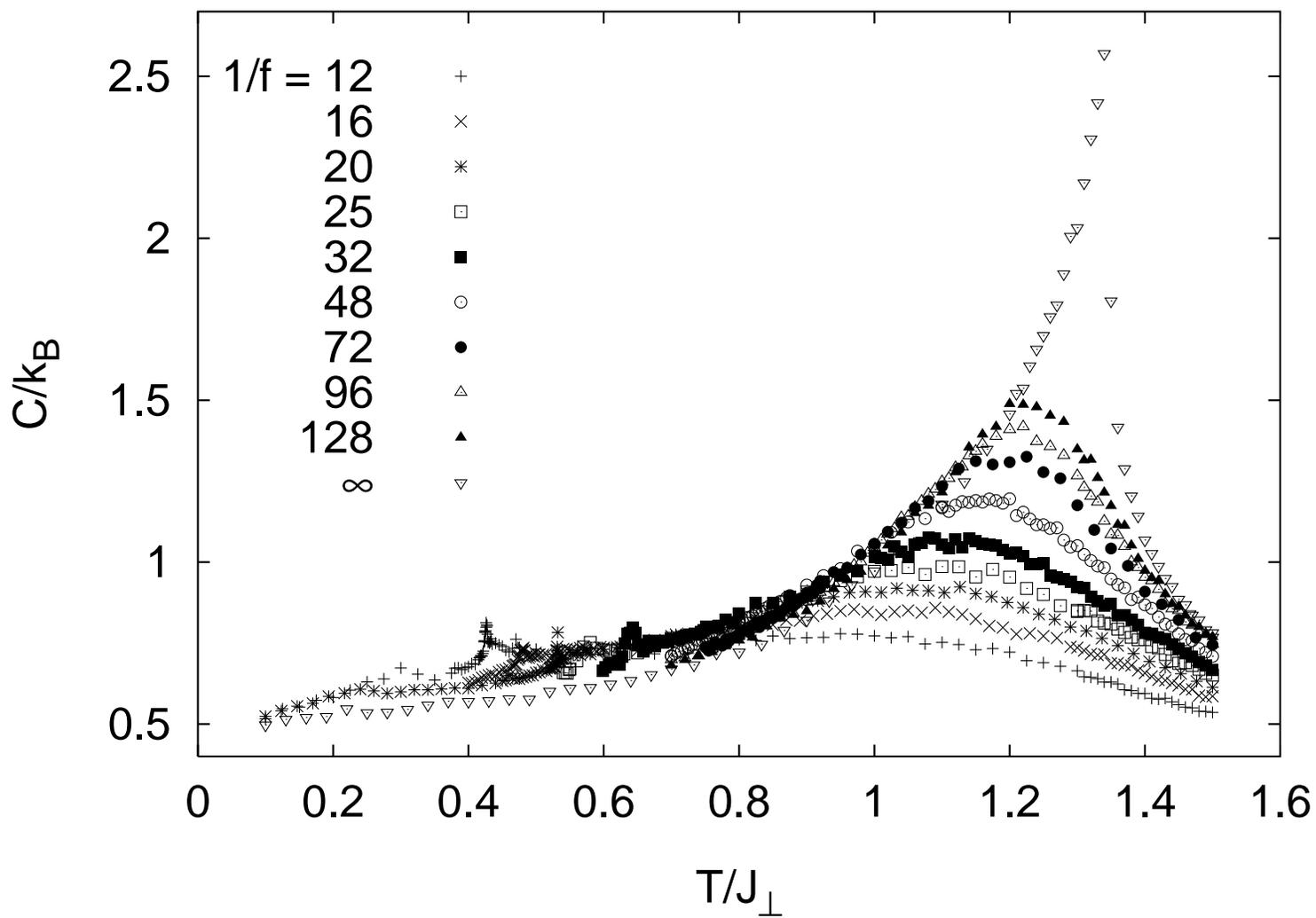
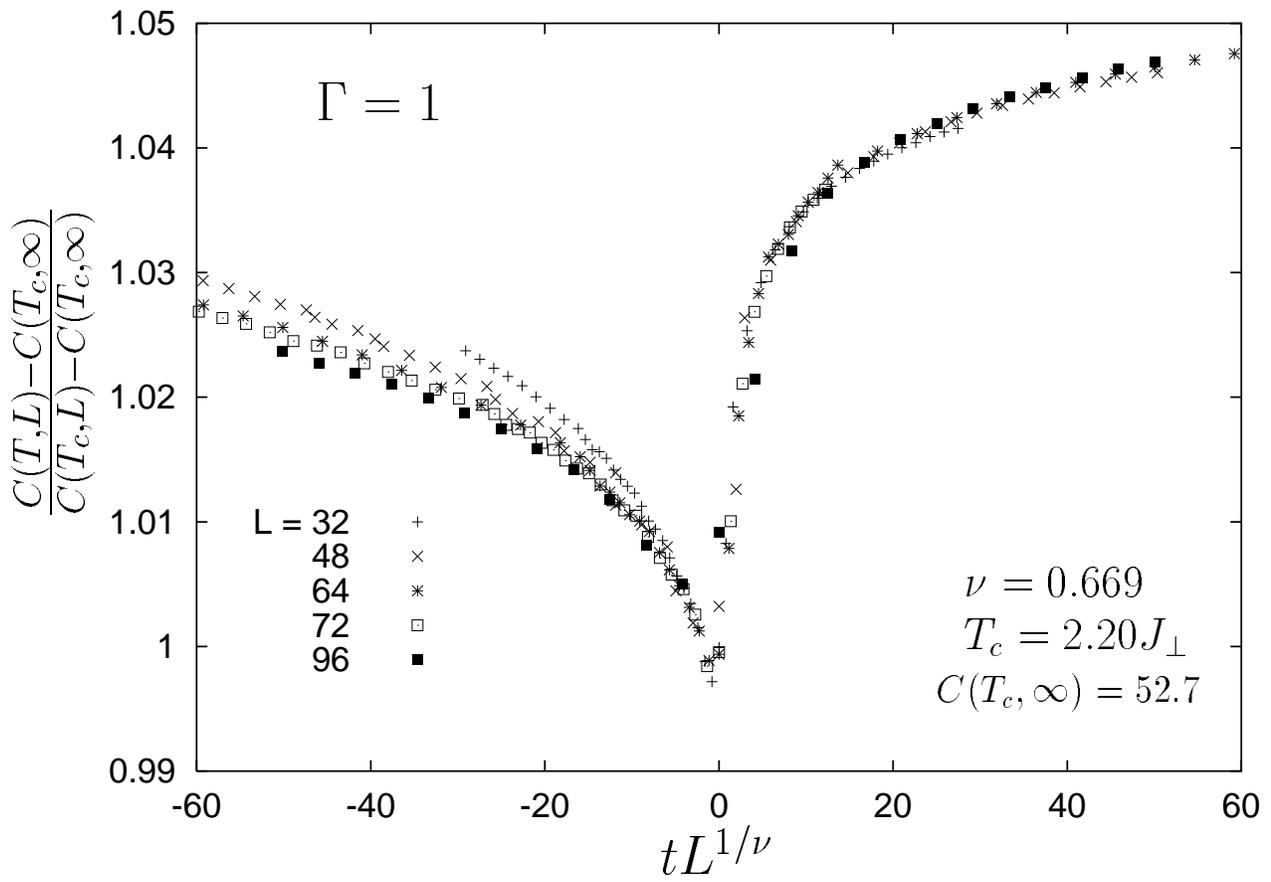
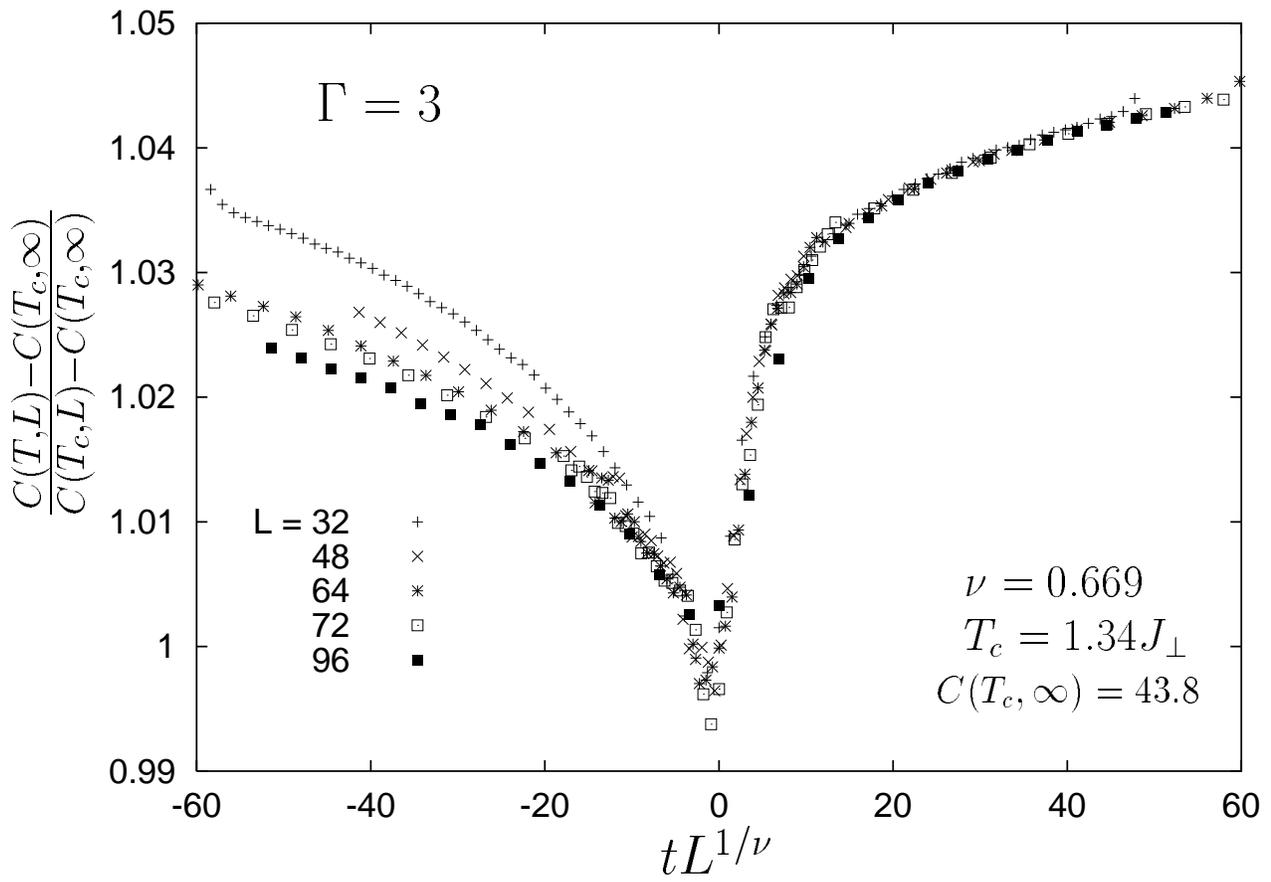
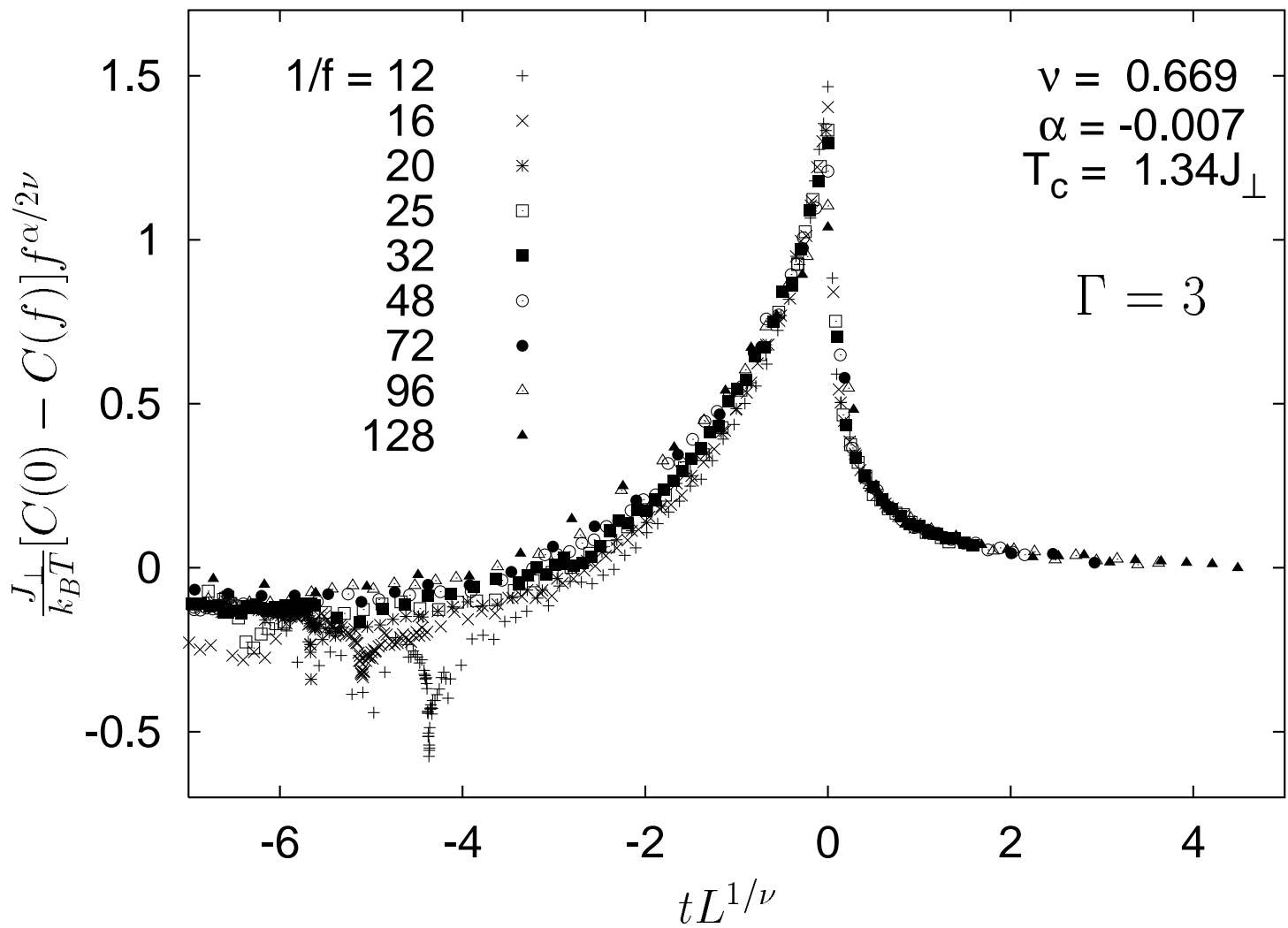
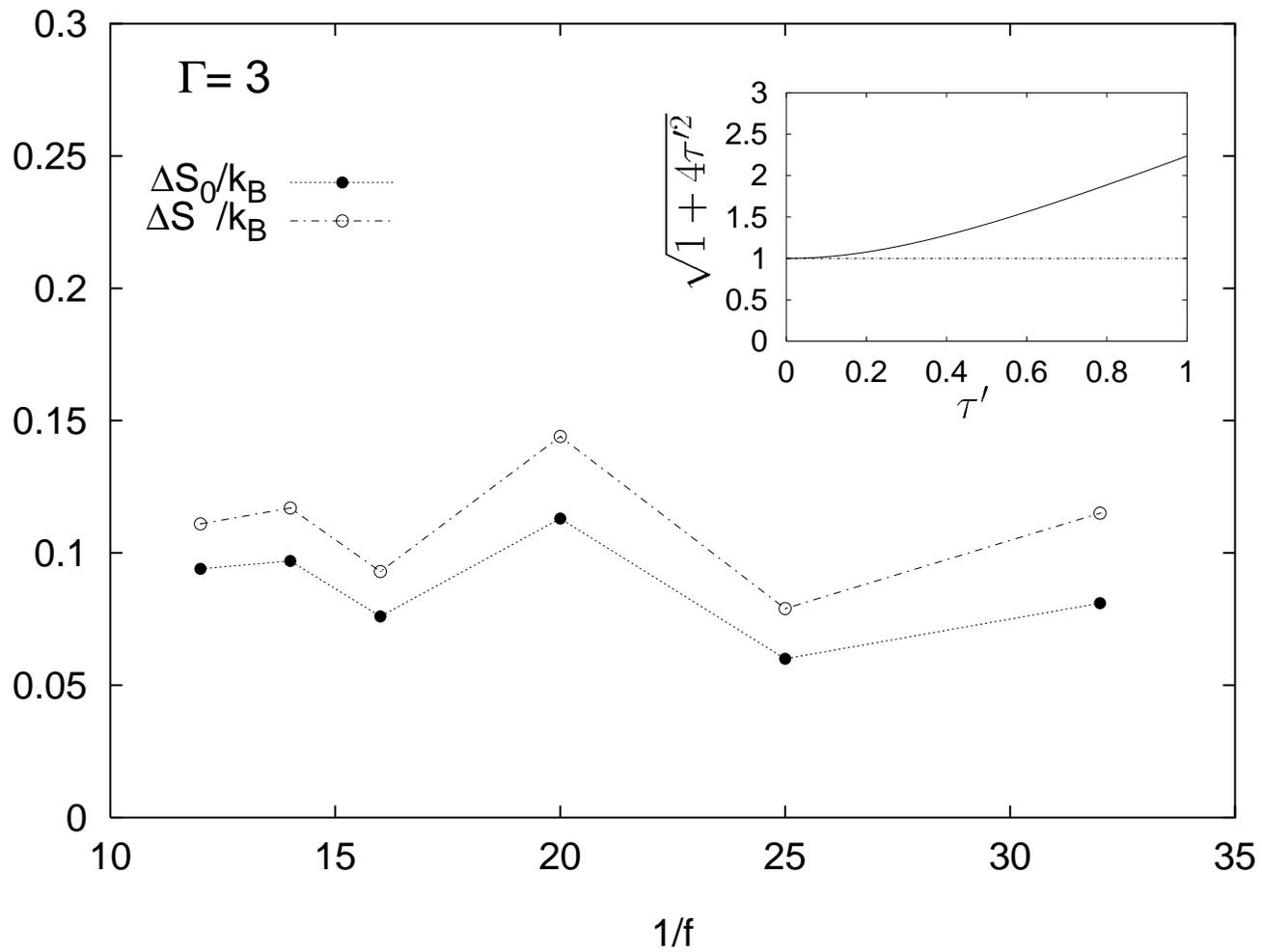
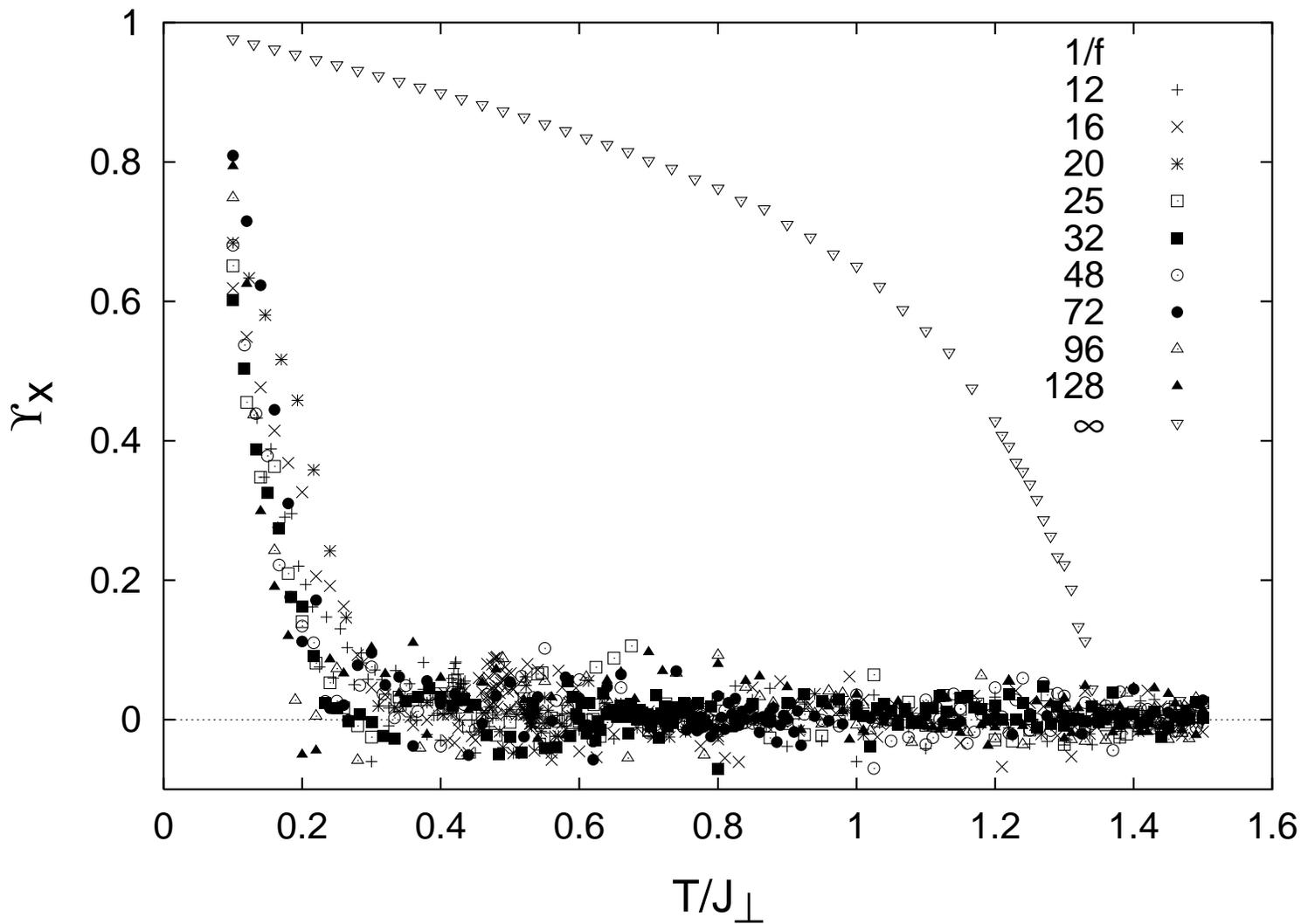
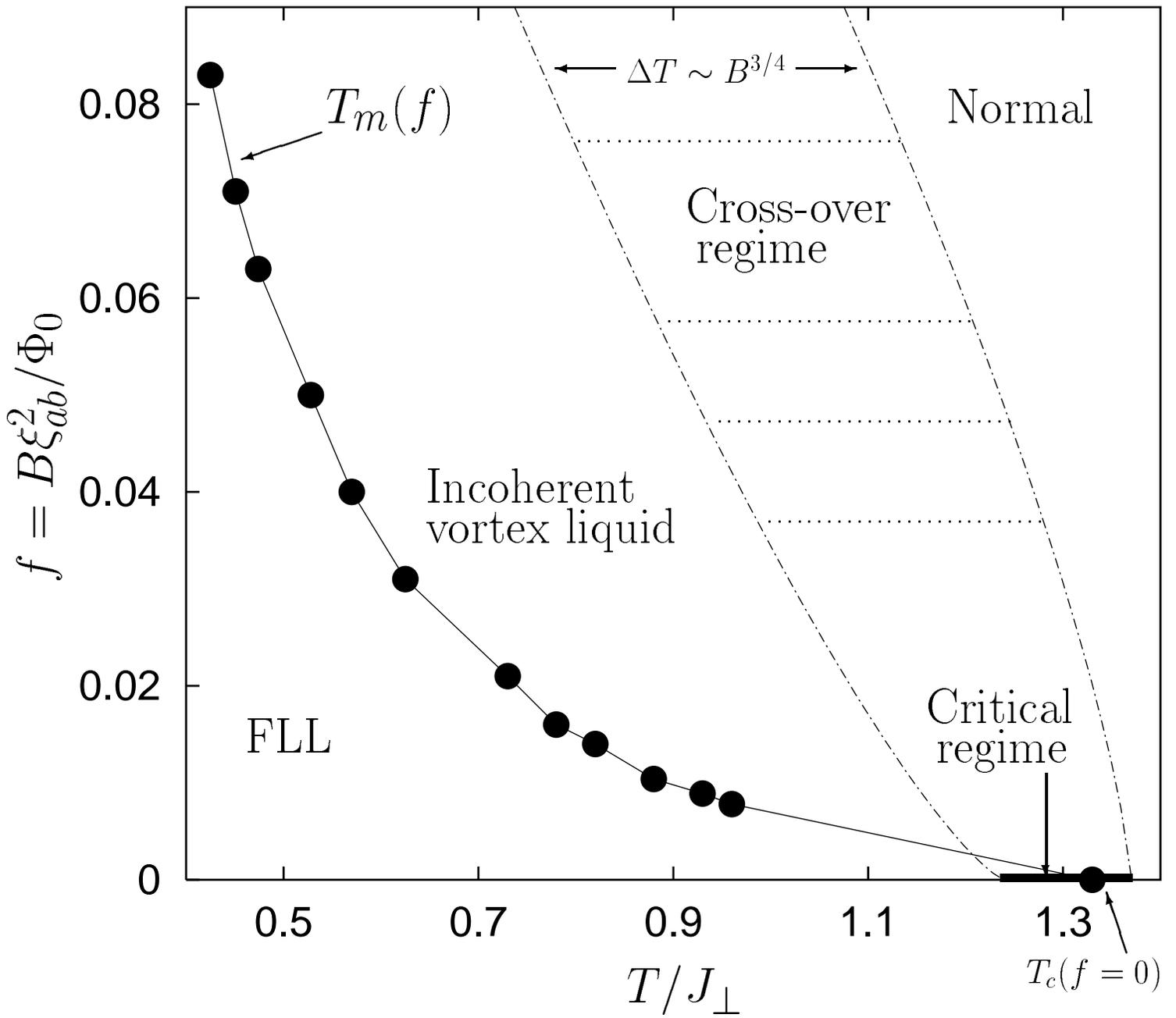
\end{document}